# Implementing Parallel Quick Sort Algorithm on OTIS Hyper Hexa-Cell (OHHC) Interconnection Network

Esam Nsour, Mohammad Fasha

## Contents

# List of Tables

# List of Figures

## Abstract

This work explores the characteristics of implementing parallel Quick Sort algorithm over the OTIS Hyper Hexa-Cell interconnection network (OHHC). OHHC interconnection architecture offers efficient processor connectivity by utilizing both electronic and optical based connections. The work presented includes analytical evaluation of the algorithm as well as simulated evaluation over a multi-threading environment. Different experiments were performed using different (OHHC) dimensions, different integer array types (random, sorted…etc.) and different array sizes. The evaluation and simulation demonstrated encouraging results that proposes the (OHHC) connectivity networks as a promising architecture. Results showed improvement in relative speedup up to 20% for OHHC full group and OHHC half group. Efficiency improvement reached up to 40% for the OHHC full group and 30% for the OHHC half group.

## 1. Introduction

In this work, we evaluate a parallel implementation of one of the well-known sorting algorithms; i.e. the Quick Sort algorithm, on the (OTIS) Hyper Hexa-Cell interconnection network (OHHC)[1]. The evaluation will include applying input data with different distribution types where each is covered by different sizes. Evaluation will be presented both analytically and by simulation.

Parallel processing or computing is a natural response for the need of speed of calculations and to process huge amount of information. Parallelism in computer science can be achieved by connecting multiple computers or processors together, or by using multiple threads on the same CPU with multiple cores. Having parallel running programs will speed up the processing time, however, it is a non-trivial thing to write parallel programs.

In this work, we will investigate the ability of parallel algorithms running on parallel architecture to produce better results than the sequential versions running on single CPU thread. In the following sub-sections we will introduce the used terminology and algorithms, and to give a description of the underlying architecture that was chosen to be simulated in our programs.

[1] https://github.com/msfasha/Algorithms/tree/main/otis_hhc_quicksort

## 1.1 Sorting

Sorting is defined as arranging data consisting of items of the same kind in a prescribed sequence. Sorting is a fundamental process which is considered as an elementary operation in computer science. Sorting is needed in too many cases where the sequence of input data is of high importance and is used as an initial step for different purpose algorithms [1].

## 1.2 Quick Sort Algorithm,

A well-known sorting algorithm first introduced by C.A.R. Hoare in 1962, sorts in place a set of data elements using the divide-and-conquer process. An array A [*p..r*] is partitioned into two non-empty sub arrays; A[*p..q*] and A[*q+1..r*]. Sub arrays are then recursively sorted by calls to Quick Sort, no combining step is required since the two sub arrays form an already-sorted array [2].

Quick sort consists of two parts, the introductory step which partitions the array into two sub arrays, and a two recursive calls for each of the resulting sub arrays. For splitting the array, Quick sort picks from the array an element called the pivot, at each step the pivot serves as a splitting point, all elements less than the pivot appear to the left side of the pivot while elements greater than the pivot appear to the right side. In all subsequent iterations, the position of this pivot will remain unchanged. When the recursion reaches the smallest size of the array; i.e. two neighboring array elements (or only one element due to bad partitioning), the recursion stops and a decision of whether to swap them (or not) is taken according to their values and the sorting direction; ascending or descending.

## 1.3 Interconnection networks

An interconnection network connects a group of processors and memory units to construct either a shared address space computers or message passing computers; it may be classified as being static or dynamic networks. An interconnection network can be represented as an undirected graph G(V,E) where a processor is represented as a node u $\in$ V (G), an edge (u, v) $\in$ E(G) between corresponding nodes u and v represents a communication channel between processors. These networks importance rise from its being the heart of the parallel processing systems. In the literature there were many topologies introduced such as Ring, Mesh, Hyper Cube…etc.

## 1.4 The Hyper-Hexa Cell (HHC)

static interconnection network is one of the topologies introduced in the literature, it consists of six connected processors arranged in hexagon shape. The hexa-shape is arranged into two triangles where each contains three processors that are fully connected, moreover, each node in each triangle is connected to the node corresponding to (facing) it in the other triangle as in figure 1.1, these six-processors grouped in two triangles represents (forms) a one-dimensional HHC. A dh-dimensional HHC is built by replacing each zero-dimension in a (dh-1)-dimensional hyper-cube interconnection network by a one-dimensional HHC undirected graph as shown in figure 1.2 [3].

## 1.5 Optical transpose interconnection system (OTIS)

An interconnection network with hybrid properties; it uses optical and electrical communication links between processors to benefit from the advantages of both types of links. Electrical links are used for neighbor processors which have short distance between them, while processors that are apart and its communication links are distanced will get an optical links in order to benefit from its speed. This type of hybrid interconnection network is referred to by “optoelectronic architectures” [3].

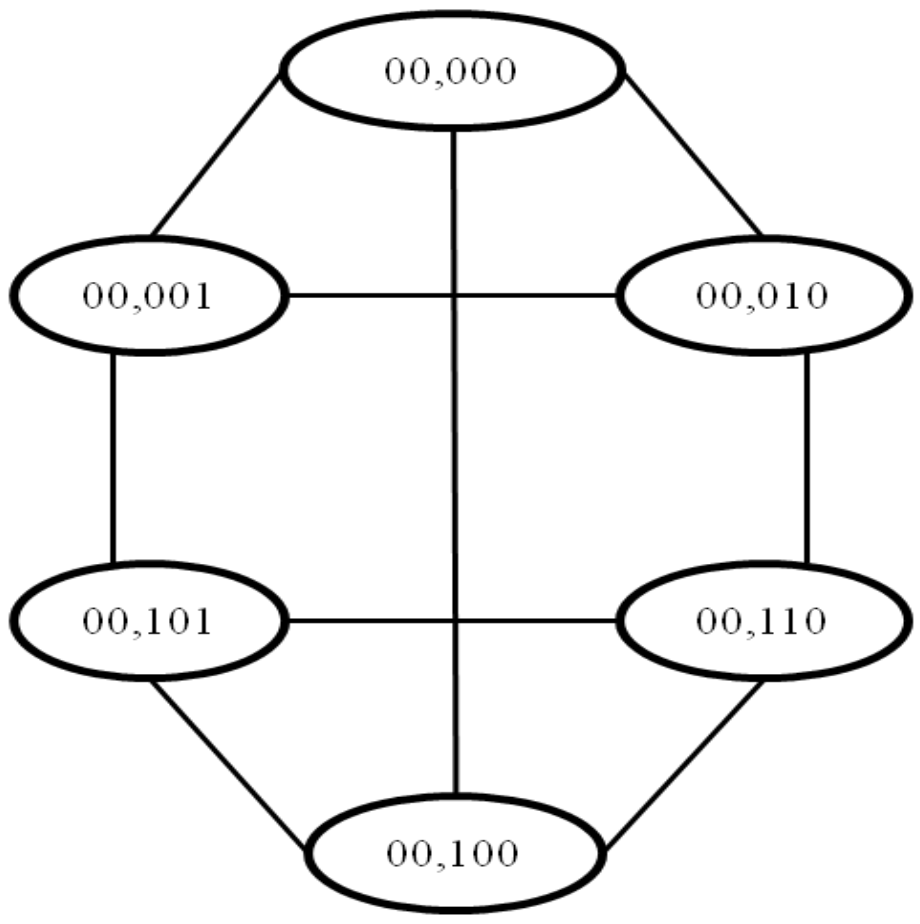

*Figure 1.1, One-dimensional HHC.*

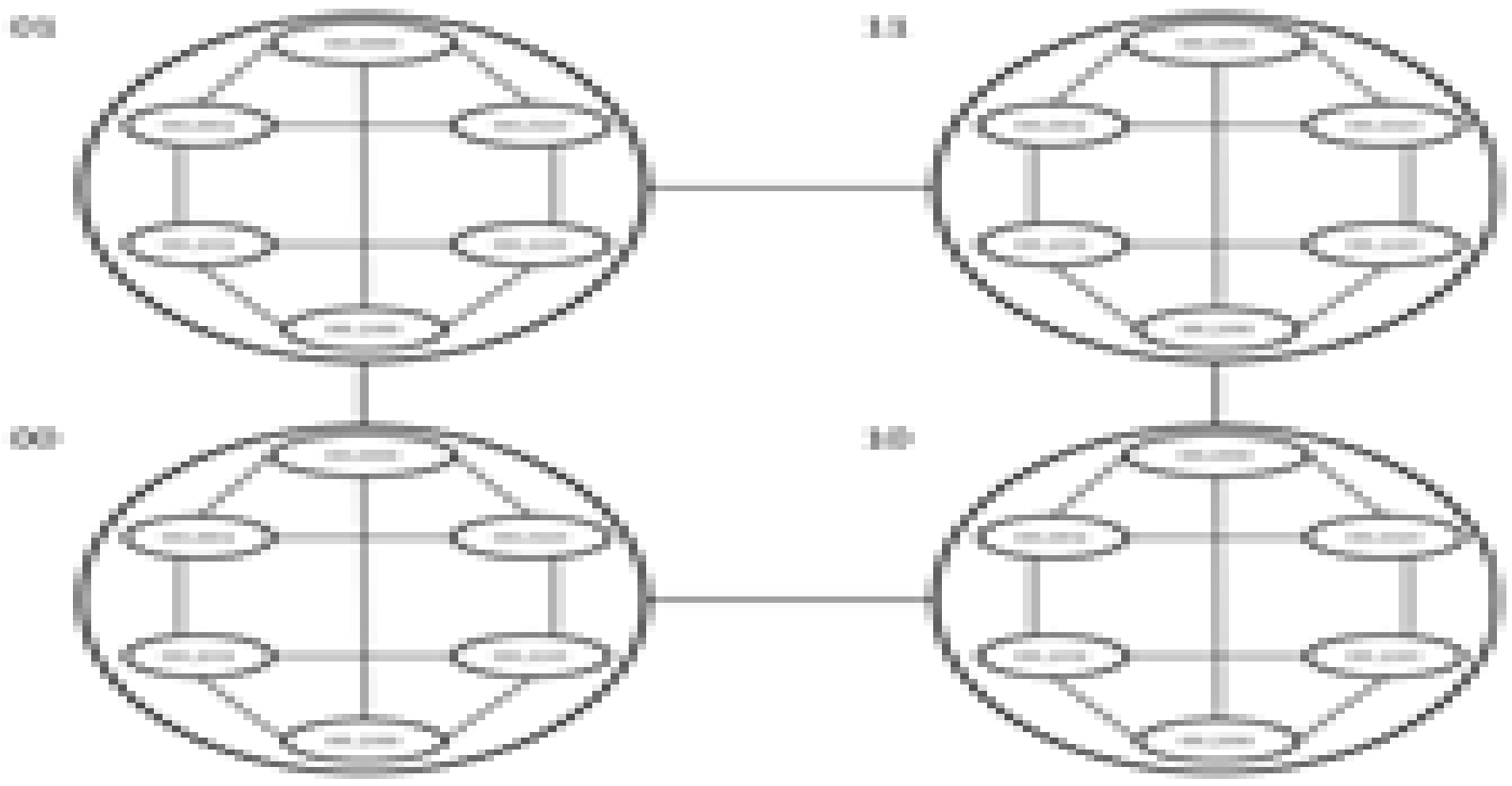

*Figure 1.2, Each zero dimension in a 2-dimensional hyper cube is replaced by one-dimensional HHC to form 3-dimensional HHC.*

OTIS Hyper Hexa-Cell (OHHC) is an optoelectronic architecture that connects a group of HHC using optical communication links between groups while keeping electrical communication links inside each group [3]. Number of groups in the (OHHC) is determined by the number of processors in the HHC. There are two methods to construct (OHHC); when the number of groups *G* is equal to the number of processors *P* at each group (the HHC dimension); i.e. *G = P*, or when number of groups *G* is equal to half number of processors *P* at each group; i.e. *G = P/2*. Figure 1.3 shows an example of the OHHC when G=P, Figure 1.4 shows the same OHHC when G=P/2 [3]. The number of processors in each dimension for the OHHC groups from the first upto the fourth dimension is given in table 1.1.

Table 1.1, The two cases of the OHHC dimensions and its corresponding number of processors.

| OHHC Dimension | When G=P | | When G=P/2 | |
|---|---|---|---|---|
| | # of groups | # of processors. | # of groups | # of processors. |
| 1-D | 6 | 36 | 3 | 18 |
| 2-D | 12 | 144 | 6 | 72, |
| 3-D | 24 | 576, | 12 | 288, |
| 4-D | 48 | 2304. | 24 | 1152. |

### 1.6 Parallel computing

Parallel computing carries out calculations using multiple processing elements simultaneously. This is accomplished by breaking the problem into independent parts so that each processing element can execute its part of the algorithm simultaneously with the others, operating on the principle that large problems can often be divided into smaller ones, which are then solved concurrently ("in parallel"). The development of parallel architecture is encouraged by the evolving computational speed and power along with decreasing costs of memory and processor components. In such architecture, large number of nodes (memories or processors) needs to communicate with one another. To support these communication requirements, many types of interconnection networks which provide a range of choices on the cost/performance spectrum have been investigated. [4]

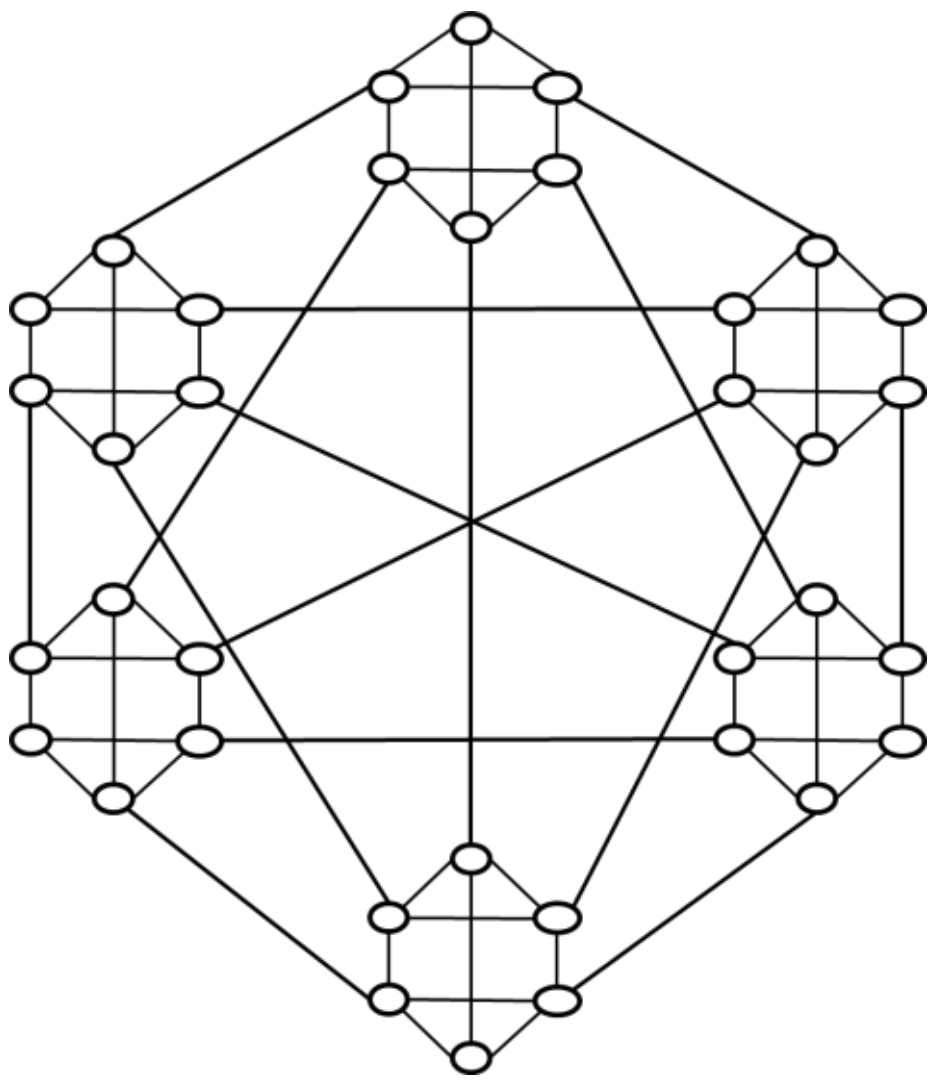

*Figure 1.3, One dimensional OHHC optoelectronic interconnection network when G=P.*

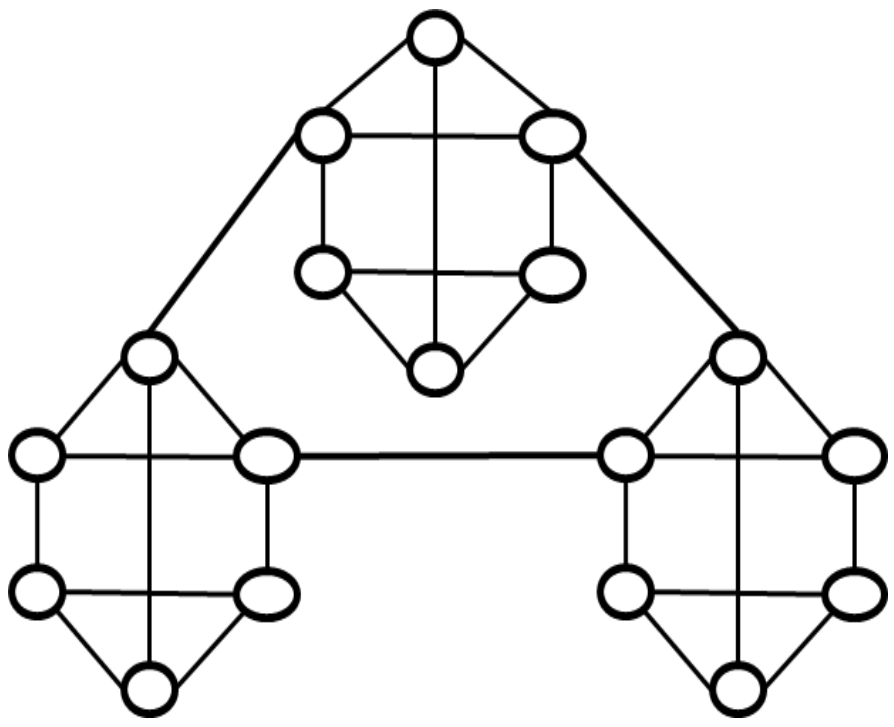

*Figure 1.4, One dimensional OHHC optoelectronic interconnection network when G=P/2.*

## 2. Related Work

The research on parallel or multithreaded versions of sorting algorithm has gained an increasing share in the last three decades motivated by the hardware improvements. Many versions of parallel and multithreaded Quick Sort versions were introduced either for general or specific architecture type. The work in [5-7] introduced Quick Sort parallelism through multithreading, while in [8] introduced the parallel Quick Sort on specific hardware (SUN Enterprise 10000), and in [9] on N-processor PRAM (parallel random-access machine).

In the literature there are so many versions of the parallel Quick Sort algorithm, but all share the same basic feature of the Quick Sort that use divide and conquer. Differences between versions are related to how and where to pick the pivots, and how and when to divide the input data into sub-divisions. The work in [10] proposed dividing the array before finding the pivots and after sorting each sub array, and to use the mean of entries in each sub array to decide where to put it in the algorithm merging phase. In [6] pivots are pre-determined in the initial phase for all needed number of sub arrays and the dividing is done according to these pivots list.

## 3. The Proposed Parallel Quick Sort Algorithm

### 3.1. Array Division Procedure

A simple approach was used to initiate payloads (sub arrays) for the designated threads (processors). The requirement was to generate a number of sub arrays according to the number of the available (given) threads. These sub arrays should be correctly divided so that the accumulated data will be automatically sorted without the need for additional arrangements in the data merge phase. To do so, the procedure creates a number of pivots (step points) that allows it to determine to which sub array an element should be sent to. This step point is determined by the following equation:

*SubDivider = (max masterArray value – min masterArray value) / (Number of Processors)*

*targetArray = masterArray[i] / SubDivider*

## 3.2. Algorithm Flow

The experimented algorithm flows as the following:

1. The user provides input defining the integer array size and type.
2. The input array is split into a number of sub arrays according to the number of the available processors "*according to the procedure explained in the previous section*."
3. The algorithm starts creating the threads and assigns them their designated sub-arrays (vectors).
4. The threads start their work by sorting their designated sub arrays.
5. Once a thread (processor) finishes sorting its designated sub array, it starts the message-passing role. During this process, each node in the architecture performs a wait and send; it waits for an amount of the payload that is statically determined by its location in the architecture.

   Each thread performs messages passing according to three general steps:

   a. The thread performs inner (HHC) data accumulation (Figure 3.1). (HHC) node (5) sends its payload directly to (HHC) node (0) and in the same time nodes (3 and 4) send their payload to nodes (1 and 2) consequently. Once node (1 and 2) receive nodes (3 and 4) payload, they pass their accumulated payloads to the designated (HHC) node (0).
   b. After (HHC) node (0) in a given (OTIS) node receives the designated (HCC) nodes payload, it starts Hyper Cube based accumulation process (Figure 3.2). In this phase, (HCC) nodes (0) across the Hyper Cube dimensions start passing payloads to their designated Hyper Cube related nodes until all the (OTIS) payloads are accumulated in (HHC) node (0) of Hyper Cube numbered (0).
   c. The last step in the data accumulation process is performed at the (OTIS) connectivity level Figure (3.3). Unlike the previous two steps which were performed using electronic-based connections, (OHHC) nodes use the optical connections between the groups to exchange messages. According to (OHHC) connectivity, node (x) in group (y) in the (OHHC) architecture is connected to node (y) in group (x). Therefore, each node passes its accumulated payload to node (y) group (0) in (OTIS) architecture. Accordingly, each head node (HHC node 0) in a given (OTIS) group passes its accumulated payload to the corresponding node in (OTIS) group (0).
   d. Finally, (HHC) nodes in (OTIS) group (0) aggregates their payloads in node (0) of the (OTIS) group according to the same procedure presented in the aforementioned steps (a and b). The only different is in calculating the wait amount which are presented in figure (3, 4) and Figure (3,5) accordingly.

```
AggregateHHC(Processor * processor)
if (HHCNodeId == 0)
    WaitForSubArrays(6);
if ((HHCNodeId == 1) or (HHCNodeId == 2))
    WaitForSubArrays(2);
    SendBackSubArrays(HHCNode(0));
if (HHCNodeId == 3)
    SendBackSubArrays(HHCNode(1));
if (HHCNodeId == 4)
    SendBackSubArrays(HHCNode(2));
if (HHCNodeId == 5)
    SendBackSubArrays(HHCNode(0));
```

*Figure 3.1, Communication step (a), Inner HHC data accumulation phase performed by all HHC Nodes except HHC Nodes in HHC Group Zero which have different rules to calculate their share of load.*

```
AggregateHyperCube(Processor * _processor)

///No need to accumulate, already aggregated in HHCNode 0 of every HHC ///Group
if (OTISDimension == 1)
return 0;

///Zero Id HHCGroups receive only, also HHCNodes > zero don't participate, only ///HHCNodeId == 0
if ((HyperCubeNodeId == 0) or (HHCNodeId > 0))
return 0;

myFirstSetBit ← GetMyFirstLeastSignificantBit();
hyperCubeDimension ← OTISDimension - 1;
waitForSubArrays ← 6 * pow(2, myFirstSetBit - 1);
sendToHHC ← HyperCubeNodeId - pow(2, myFirstSetBit -1);

WaitForSubArrays(waitForSubArrays);
SendBackSubArrays(sendToHHC);
```

*Figure 3.2, Communication Step (b), Hyper Cube data accumulation phase performed by all HHC Groups except HHC Group zero which has a different rule to calculate its share of load.*

```
if ((OTISNodeId == 0) and  (OTISGroupId != 0))
{
///make sure that you've received the correct load, otherwise, wait
WaitForSubArrays ((6 * pow(2, OTISDimension-1));

///resolve the designated OTIS node in OTIS group 0
targetHHCGroupId ←  ((OTISGroupId / 6) + 1) - 1;
targetHHCNodeId ←  OTISGroupId % 6;
SendTo ← OTISGroupId * (6 * pow(2,OTISDimension -1)) + 6 * targetHHCGroupId + targetHHCNodeId
SendBackSubArrays(SendTo);
}
```

*Figure 3.3, Communication Step (c), OTIS data accumulation performed by each OTISNode Zero except OTISGroup Zero which has a different rule to calculate its share of load.*

```
normalHHCNodeWaitFor = GetHyperCubeNodesNumber((OTISDimension) * 6) + 1;
aggregateHHCNodeWaitFor = normalHHCNodeWaitFor * 2;
normalHHCHeadNodeWaitFor = normalHHCNodeWaitFor * 6;
masterHHCHeadNodeWaitFor = (normalHHCNodeWaitFor * 5) + 1;

if (HHCNodeId == 0) and (HHCGroupId == 0))
    WaitForSubArrays(masterHHCHeadNodeWaitFor);
else if (HHCNodeId == 0)
    WaitForSubArrays(normalHHCHeadNodeWaitFor);

if ((HHCNodeId == 1) or (HHCNodeId == 2))
{
    WaitForSubArrays(aggregateHHCNodeWaitFor);
    SendBackSubArrays(OTISGroupId,HHCGroupId, 0);
}

if (HHCNodeId == 3)
{
    WaitForSubArrays(normalHHCNodeWaitFor);
    SendBackSubArrays(OTISGroupId,HHCGroupId, 1);
}

if (HHCNodeId == 4)
{
    WaitForSubArrays(normalHHCNodeWaitFor);
    SendBackSubArrays(OTISGroupId,HHCGroupId, 2);
}

if (_processor->HHCNodeId == 5)
{
    WaitForSubArrays(normalHHCNodeWaitFor);
    SendBackSubArrays(OTISGroupId,HHCGroupId, 0);
}
```

*Figure 3.4 presents the wait for rules for OTIS group 0, the flow is the same as Figure (3.1) while the difference is in calculating the static amount of sub-arrays that OTIS group 0 HHC nodes should wait for in the inner HHC phase.*

```
normalHHCNodeWaitFor = (GetHHCGroupsNumber(OTISDimension) * 6) + 1;
normalHHCHeadNodeWaitFor = normalHHCNodeWaitFor * 6;
waitForSubArrays = normalHHCHeadNodeWaitFor * pow(2, mySetBit - 1);
```

*Figure 3.5, The HyperCube accumulation step at OTIS group 0, the flow of the algorithm is similar to conventional accumulation by other groups (Figure 3.2) while the difference is in the wait for calculation.*

## 4. Analytical assessment

The presented parallel Quick Sort algorithm properties were analyzed analytically. Algorithm Properties are directly related to the features of the underlying architecture (OHHC); i.e. algorithm was designed to simulate it. The following sub sections present major algorithm properties.

### 4.1. Time Complexity

Complexity represents time spent by a given algorithm to complete its work measured by the number of steps needed. Quick Sort is considered one of the fastest sorting algorithm, for its sequential version worst case; at each partitioning step there are one of the resulting sub arrays size is equal to one, it has a running time complexity of $\Theta(n^2)$, but in general, the expected average and best sequential running time is $\Theta(n \log n)$. The best case in sequential Quick Sort occurs when the array is always partitioned into two equal parts in every step. For the parallel version of the Quick Sort, there are additional steps that are related to sending or receiving data from other processors, or creating and destroying threads, which are the steps that make parallel sorting feasible. However, time complexity calculation of the parallel version presented in this report will overlook these steps. Average parallel time complexity is given by theorem 1 below.

***Theorem 1:*** *The average time complexity of the parallel version of the Quick Sort algorithm is* $\Theta(n/P \log n/P)$.

Proof: The original input array of size $n$ is partitioned into $t$ chunks that represent each processor share of the original array; $t = n/P$, where $P$ is number of processors. Each processor will sort its array chunk $t$ using the sequential version of the algorithm which takes $\Theta(t \log t)$ in parallel with all processors. Assuming that all processors get almost the same share $(t)$, and not considering the splitting ,distributing, and gathering of the array chunks then then total time complexity will be $\Theta(n/P \log n/P)$.

### 4.2. Number of Communication Steps

Communication Steps represent the number of steps required by algorithm to spread the array chunks from the head node to its destination at a specific processor and steps for sending these chunks back to the head processor after being sorted. Nodes in the interconnection network will communicate in some order dictated by the physical connections pattern between it to perform the needed steps for the algorithm to finish the sorting job. Total number of steps are the sum of the steps inside the head OHHC group and the destination OHHC group that need to go through intermediate nodes at each group in addition to the optical link between groups. The total electronic links communication steps are $12*G*d_h - 2G$ and total optical links communication steps are $2G-2$. Total number of communication steps is given in theorem 3.

***Theorem 3:*** *Total number of communication steps for the proposed parallel Quick Sort on the OHHC interconnection network from source to destination and back to source is:* $12*G*d_h -2$

Proof: During the distribution of array chunks phase, there are 5 steps inside each group (electrical links steps) at each 1-dimensional HHC; 3 steps to spread chunks from the head node to the three directly connected nodes in addition to 2 steps to spread array chunks to the other remaining two nodes. Between HHC dimensions there are $((d_h-1) * 6)$ steps. Total sum for each group (electrical links steps) is: $6*d_h-6 + 5 = 6*d_h-1$, and for all groups $= G* (6* d_h-1)) = 6*G*d_h - G$. To include the optical link steps we will use G which is the number of groups in the OHHC to represent the optical steps, since the array chunks are sent to each group head node from nodes at group zero then there are G-1 optical steps, the value of G is determined by the rule; G= P when using full OHHC structure, or G=P/2 when using the half-full OHHC structure. Adding these optical steps to total electrical steps, the total communication steps is $= (G-1)+ (6*G*d_h – G) = 6*G*d_h -1$. For the array chunks to be sent back to the head node of group zero, the same amount of steps will be needed, so the total communication steps from source node to destination nodes and back to the source node is: $2*(6*G*d_h -1)= 12*G*d_h -2$

## 4.3. Speed Up

speed Up metric measures the percentage of improvement in time complexity when using a parallel version of an algorithm by comparing it with the sequential version complexity. The ratio of improvement is measured by the equation: Speedup (S) = Serial run time ($T_S$) / Parallel run time ($T_P$). Speedup of the parallel version presented here is given in theorem 4.

***Theorem 4****: Speedup of the parallel version of Quick Sort in the average case is:* $\Theta (P \log n / ( \log n – \log P ))$

Proof: Since $T_S = \Theta(n \log n)$ and $T_P = \Theta(n/P \log n/P)$, Then:

$$T_S/T_P = \Theta((n \log n) / (n/P \log n/P))$$

$$= \Theta((P/n)*n \log n / ( \log n/P))$$

$$= \Theta(P \log n / \log n/P)$$

$$= \Theta(P \log n / ( \log n – \log P ))$$

## 4.4. Efficiency

The efficiency metric measures the ratio between the sequential run time with the parallel run time in all processors accumulated. The ratio given by the equation: Efficiency (E) = Speedup (S) / number of processors (P), and can be rewritten as E= Serial run time ($T_S$) / P * Parallel run time ($T_P$), it means that if the time taken by all processors in parallel accumulated and compared to the serial time of the algorithm it will present the effectiveness of the parallel algorithm. In the best case the total parallel time will equal to the sequential time and then the system is called ideal. The efficiency of the proposed algorithm is given in theorem 5.

***Theorem 5:*** *Efficiency of the parallel version of the Quick Sort algorithm is:* $E_f = \Theta(\log n / \log n - \log P)$.

Proof: The efficiency is equal to Speedup divided by the number of processors P; $E_P = T_S/PT_P$ . Which equals

$$\Theta((n \log n) / P^* (n/P \log n/P))$$

$$= \Theta((n \log n) / (n \log n/P))$$

$$= \Theta((\log n) / (\log n/P))$$

$$= \Theta((\log n) / (\log n - \log P))$$

## 4.5. Message Delay

This metric is presents the maximum time needed by any message to travel from source to destination, in other words it is the time for a chunk of array data to be sent from the head node to the processor that will sort it sequentially or its way back. The message delay is given in theorem 6.

***Theorem 6****: Message delay when using the parallel version of Quick Sort at the OHHC interconnection network is:* $\Theta (t + ((2 * d_h +3)$ *in the average case, and* $\Theta (n + ((2 * d_h +3)$ *in the worst case.*

Proof: The head node will prepare a unique chunk to be send to each processor, assuming the average case where all chunks are almost equal, the message size (t) will equal n/P; where n is the total array size and P is number of processors. The longest path for a message is the diameter of the source group plus the diameter of the destination group plus one optical connection link from the source group to the destination group, that is number of links (L) = ($2 * d_h +3$). Assuming we are using the store-and-forward routine, the cost of $\Theta(t * L)$ is applied and the message delay will be $\Theta (t * (2 * dh +3))$. When assuming the worst case of partitioning; the array chunks sizes will be either too small or very big and is almost equal to the original size, in this case $t \approx n$.

## 4.6 Summary of analytical assessment.

The analytical assessment is summarized in table 4.1 below.

Table 4.1, Summary of the analytical assessment of the proposed algorithm

| | |
|---|---|
| Time Complexity | $\Theta\ (n/P \log n/P)$ |
| Number of communication steps | $12*G*d_h -2$ |
| Speedup: | $\Theta(P \log n / ( \log n – \log P ))$ |
| Efficiency: | $\Theta((\log n) / (\log n – \log P))$ |
| Message delay; Worst case of partitioning ; t ≈ n | $\Theta\ (n * ((2 * d_h +3)$ |
| Message delay; Average case of partitioning; t ≈ n/P | $\Theta\ (n/P * ((2 * d_h +3)$ |

## 5. Simulation Environment

Several simulations where performed to examine the performance of parallel Quick Sort over the OTIS Hyper Hexa-Cell optoelectronic architecture. The experiments where run over the following environment:

- Hardware
  - Intel Core i 7 @ 2.2Ghz dual (quad cores) machine.
  - 6 GB Ram.
- Software
  - Windows 7 OS 64 bit version.
  - GNU C++ version 4.7 using Code Blocks IDE.

Integer arrays were used as input data to the algorithm and a combination of the following variances where experimented:

a. Using different types of integer arrays (random generated arrays, sorted arrays, reverse sorted arrays, local distribution version of the input array).
b. Using different array sizes (10, 20, 30, 40, 50 and 60) MB arrays.

Initially, the algorithm was examined using a conventional sequential approach to capture the sequential algorithm performance while running over all the aforementioned combinations. This was important to provide baselines that can be used to analysis and comparison with the corresponding parallel runs.

Later, the same algorithm was experiment over the OTIS Hyper Hexa-Cell (OHHC) architecture. For this purpose, two main architectures of (OHHC) were examined:

*a.* The full OTIS Hyper Hexa-Cell optoelectronic architecture where the number of (OHHC) groups is equal to the number of processors in each group.
b. A minimized version of the OTIS Hyper Hexa-Cell optoelectronic architecture where the number of (OHHC) groups is equal to half the number of the processors in each group.

Moreover, using the aforementioned two OHHC architectures, several experiments where performed on different OHHC dimensions, the 1-dimension, 2-dimension, 3-dimension and the four (OHHC) dimension.

In brief, different 216 runs where examine using different combinations of the variances that were presented in the previous section.

# 6. Experimental Results and Performance Analysis

## 6.1. Execution time

The first observation was related to the sequential version of the algorithm. In this scenario, different array sizes and different array types were experimented. In general, the increase in array sizes caused an expected linear increase in execution times. This is rational since the number of computations is increased according to array sizes. Also, it was observed that the algorithm performed better when no random values were involved i.e. sorted and reverses sorted arrays. Figure 6.1 presents a comparison of the different sequential sorting runs.

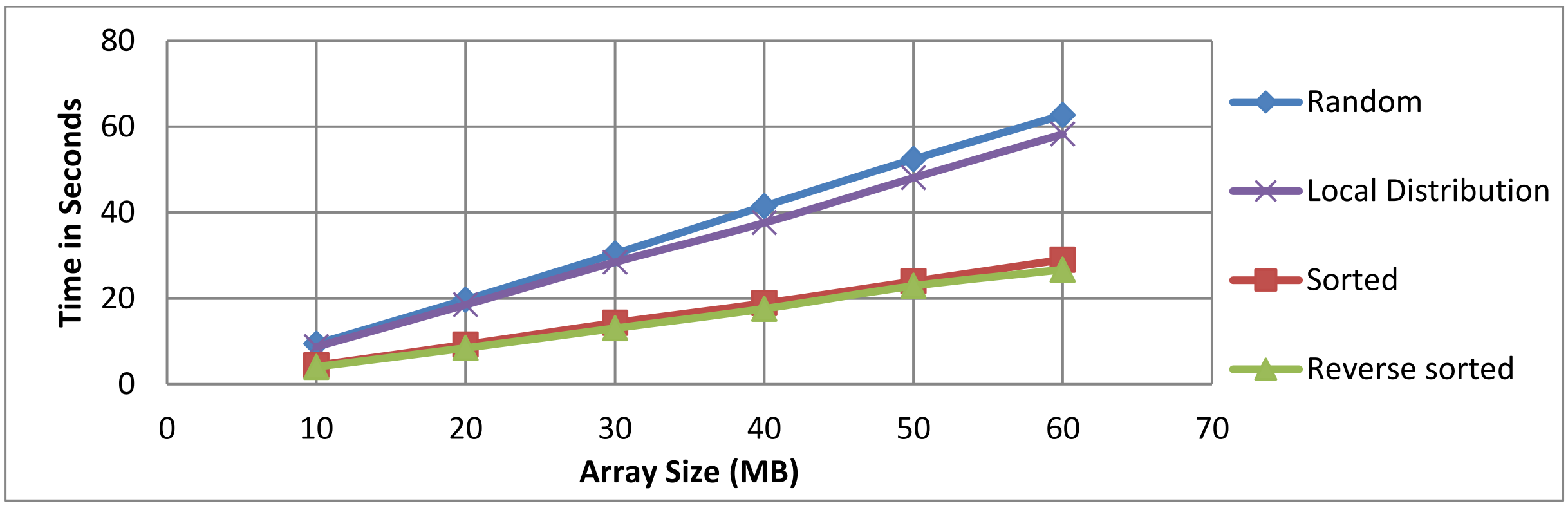

*Figure 6.1, Running the sequential version of the algorithm over arrays of different types and different sizes*

While for the parallel run of the algorithm, several combinations of different integer array sizes and types were experimented. The general flow also demonstrated a co-relation between array sizes and the execution. Also, the higher the (OHHC) dimension the better results are achieved i.e. less execution time. This can be explain because higher (OHHC) dimensions yield more processing resources, and the main array can be split to more sub arrays which will be sorted in parallel. This is compliant with the previously presented theorems where the increase in the number of processing units shortens the execution times. Moreover, using real environment dedicated processors shall be able to reach even lower times that the ones that were achieved using the multi-threaded simulation. Figure 6.2 presents the execution times when running different dimension (OHHC) network over a random array with different sizes.

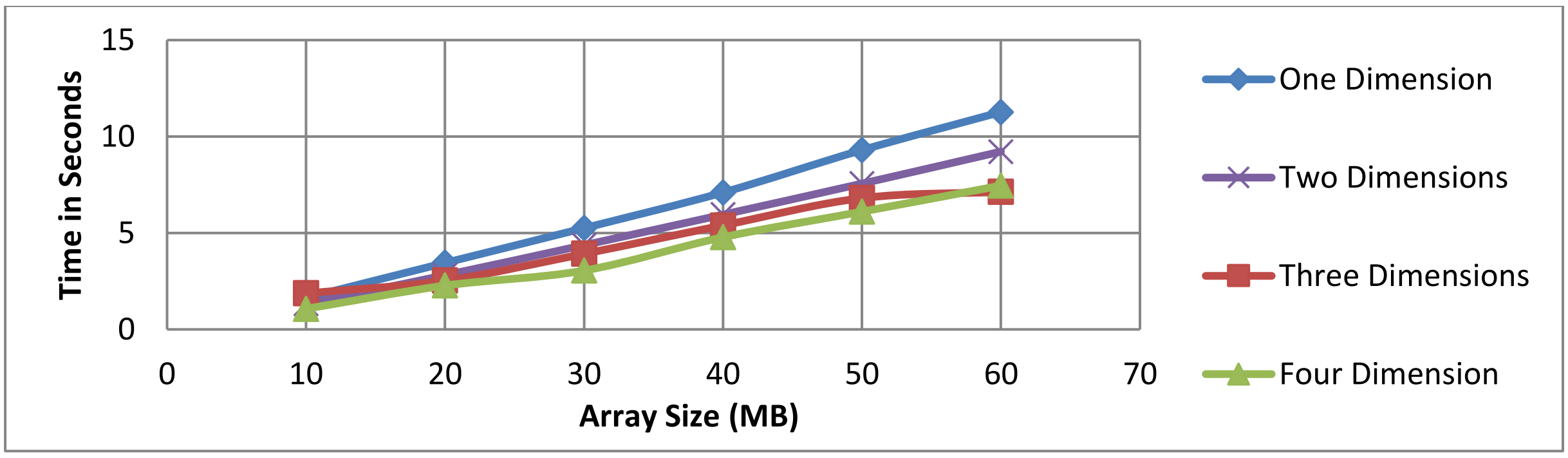

*Figure 6.2, Run time of the Parallel Quick Sort algorithm over different OHHC dimensions using Random distribution.*

Other noticeable result which is presented in the figure below is that the parallel sort algorithm demonstrated the same behavior of the conventional sequential run presented in the first paragraph. Figure 6.3 demonstrates the performance of four dimensions of (OHHC) network over integer arrays of different types and different sizes. Random-based array types yielded worse results than the non-random versions i.e. random and local distribution vs sorted and reverse sorted arrays.

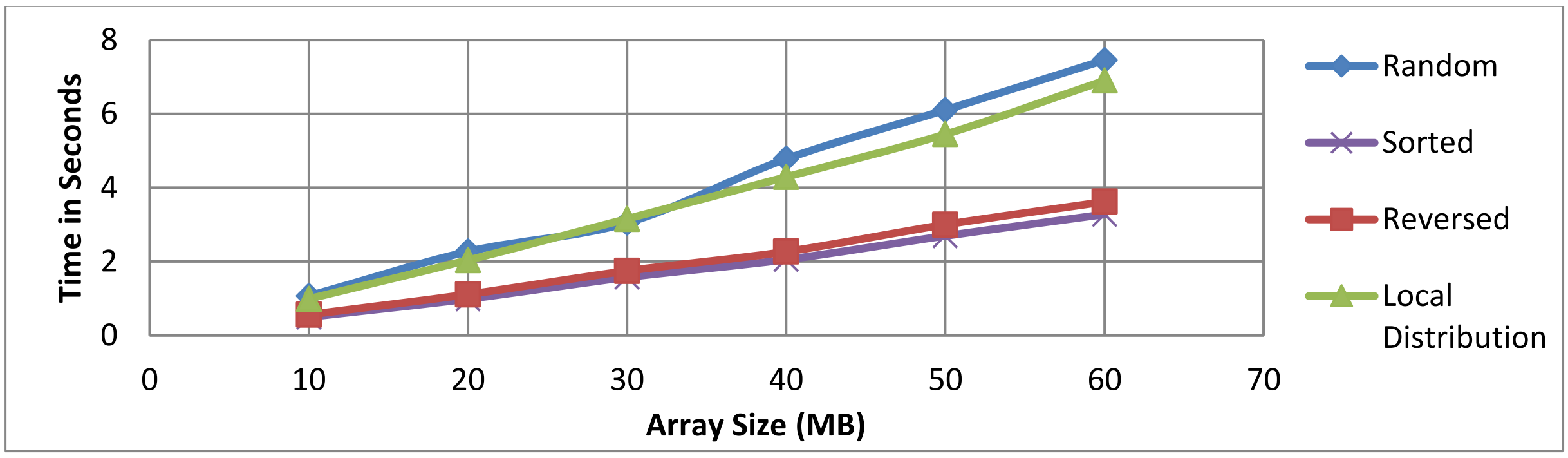

*Figure 6.3, A Parallel run using a four-dimension network sorting integer arrays of different types and different sizes*

## 6.2. Relative speed up

Relative speedup $S_P$ is the ratio between the time to solve the problem by the sequential version $T_S$ to the time of the last thread finish of the parallel version $T_P$ when sorting the same input array size and distribution type. In the following subsection we will introduce the speedup when number of processors is equal to number of groups, i.e. G = P, while in the next subsection we will introduce the speedup for the case where G = P/2.

### 6.2.1 Relative speedup when *G = P OHHC*

Figure 6.4 demonstrates the amount of improvement percentage when using random distribution in different array sizes for the parallel version over the sequential version when G=P and number of dimensions is 1, 2, 3, and 4 . Figure shows the amount of speedup is almost steady when the number of processors increase for all array sizes except for the case where dh = 3 which shows increasing speedup when array size increase. The relative speed up did not get above 10% even when a 4-d OHHC was used, this is due to the data distribution which forces the algorithm to take more time is sorting the sub arrays at each individual processor.

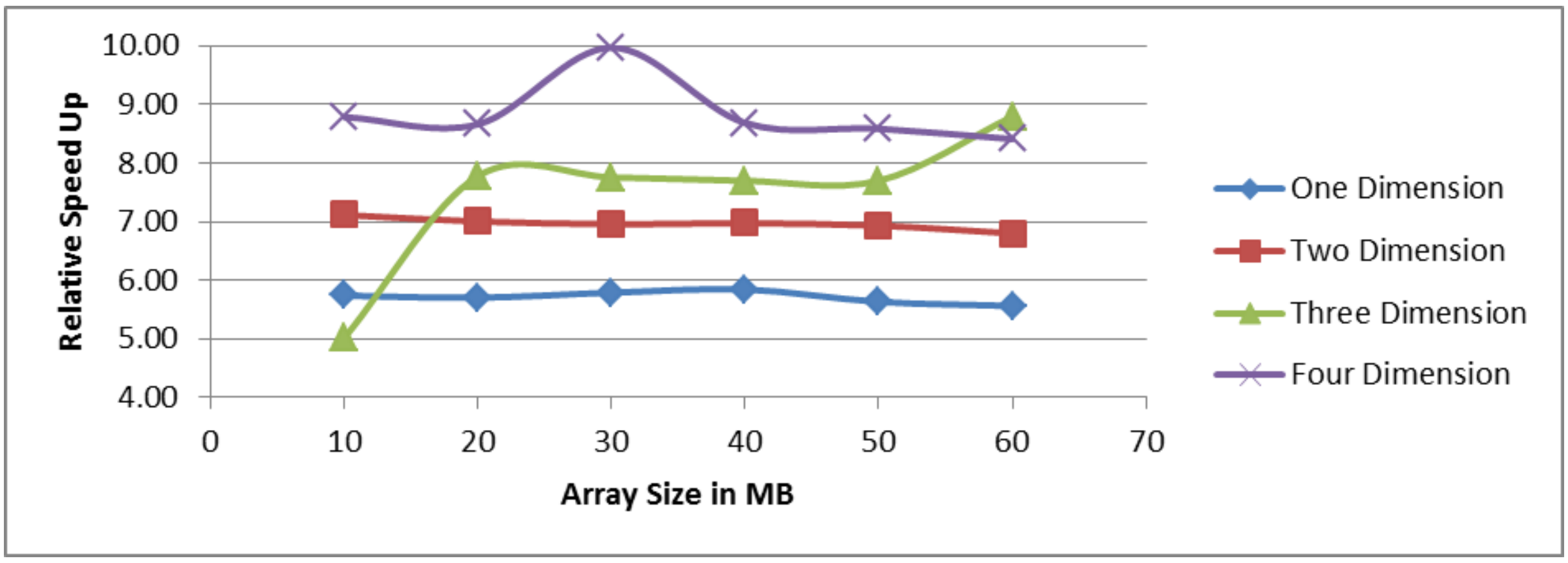

*Figure 6.4, Relative Speedup when G=P using Random distribution for different OHHC dimensions*

Relative speedup for the case where already sorted array is input to the algorithm, it shows the amount of improvement percentage in different array sizes for the parallel version over the sequential version when G=P and number of dimensions is 1, 2, 3, and 4 . Figure 6.5 shows the amount of speedup is almost steady when the number of processors increases for all array sizes. The Speed ratio shows improvement up to 20%.

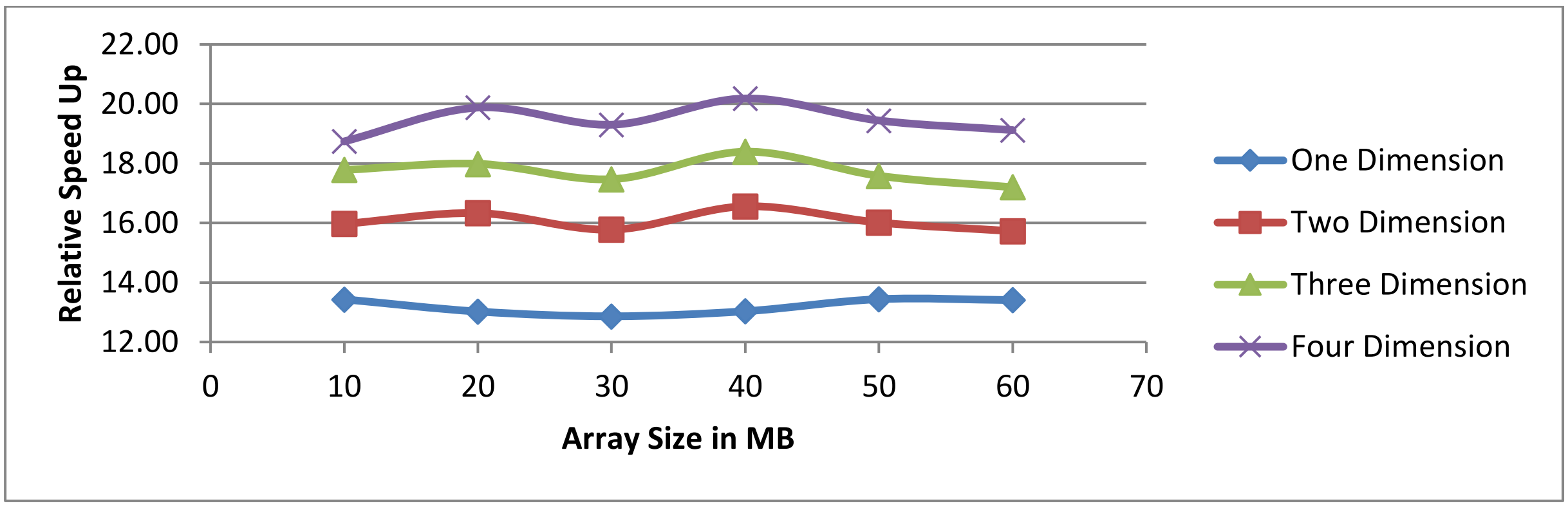

*Figure 6.5, Relative Speedup when G=P using sorted distribution for different OHHC dimensions*

Figure 6.6 shows the relative speedup for the case where Reversed sorted array was input and G=P. As the case in the Sorted array, there is a good speed up that reaches up to 18%.

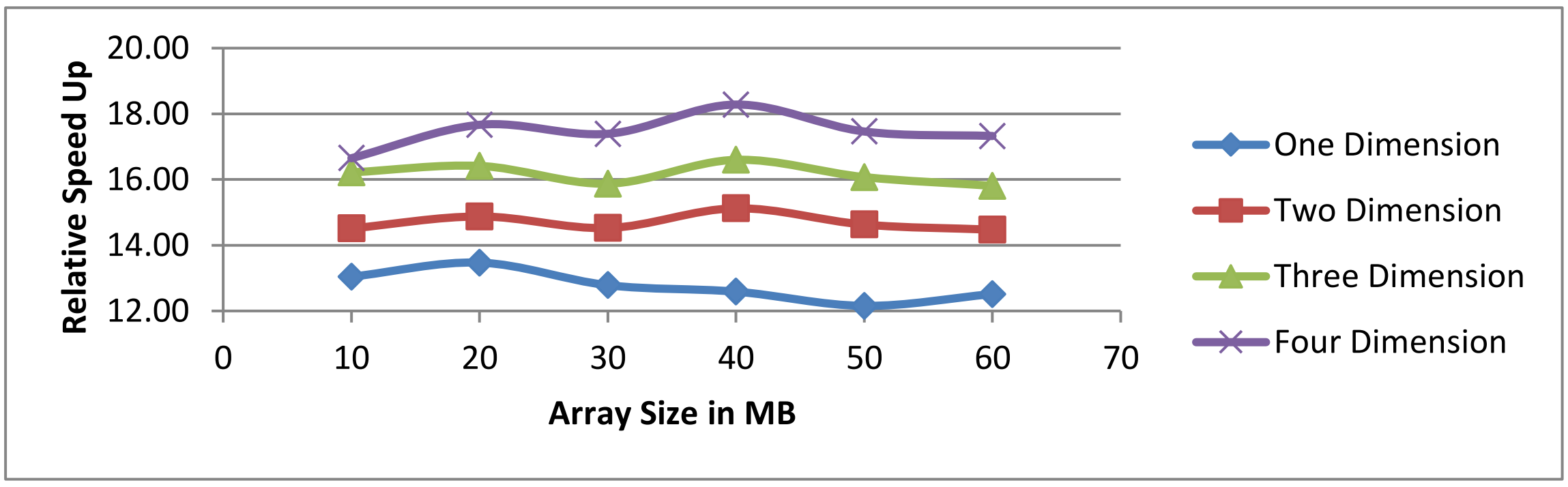

*Figure 6.6, Relative Speedup when G=P using reversed sorted distribution for different OHHC dimensions*

For the local distribution data array, relative speedup in figure 6.7, the relative speed shows no more than 10% when dimension used is 4. This is due to the same reason stated in the Random array input case.

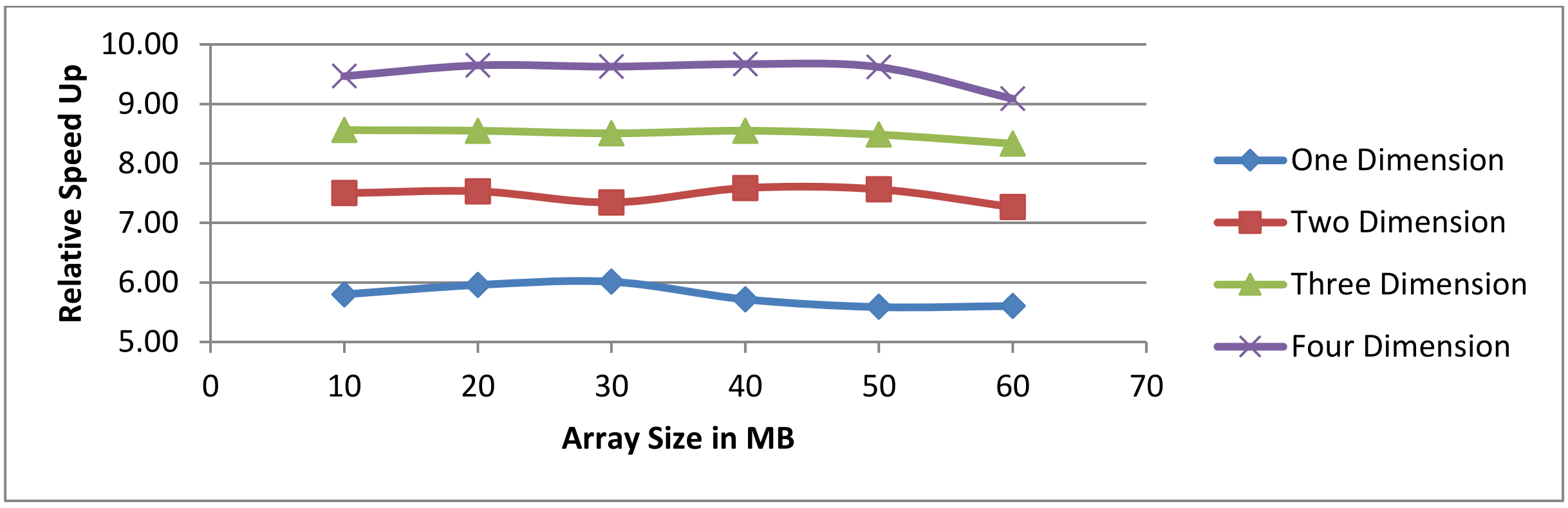

*Figure 6.7, Relative Speedup when G=P using local distribution data input for different OHHC dimensions*

### 6.2.2 Relative speedup when G = P/2 OHHC

This subsection introduces the Relative speedup when number of groups is equal to half of processors number of, i.e. G = P/2; i.e. half the number of processors in the $d_h$ HHC. Relative speedup were obtained for different distribution types and array sizes. The Figures of this subsection are representing the relative speedup for array sizes from 10MB up to 60MB of data on the x-axis and the relative speed up on the y-axis for different HHC dimensions namely from 1 to 4 dimensions.

Figure 6.8 shows the amount of improvement when using random distribution for the parallel version over the sequential version when G=P/2 and number of dimensions is 1, 2, 3, and 4 . Amount of speedup is increasing when the number of processors increase, and almost steady for all array sizes. For the case where $d_h$ = 4 we only could get results for array size less than 40MB. The maximum speed up almost reached 9%, This is due the time consumed by the nodes to sort the sub arrays is longer in this distribution type than the sorted and reversed sorted distribution types.

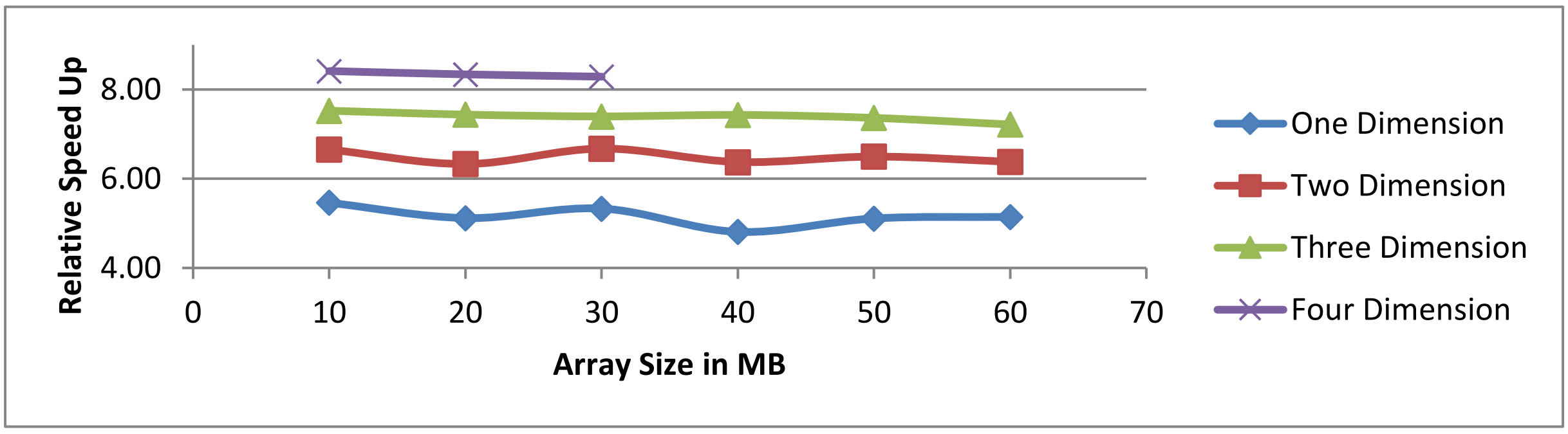

*Figure 6.8, Relative Speedup when G=P/2 using Random distribution for different OHHC dimensions.*

Figure 6.9 shows the amount of improvement when using Sorted distribution for the parallel version over the sequential version when G=P/2 and number of dimensions is 1, 2, 3, and 4 . Amount of speedup is increasing when the number of processors increase, and almost steady for all array sizes. For this type of distribution the speed up reached 20% when using 4-d OHHC.

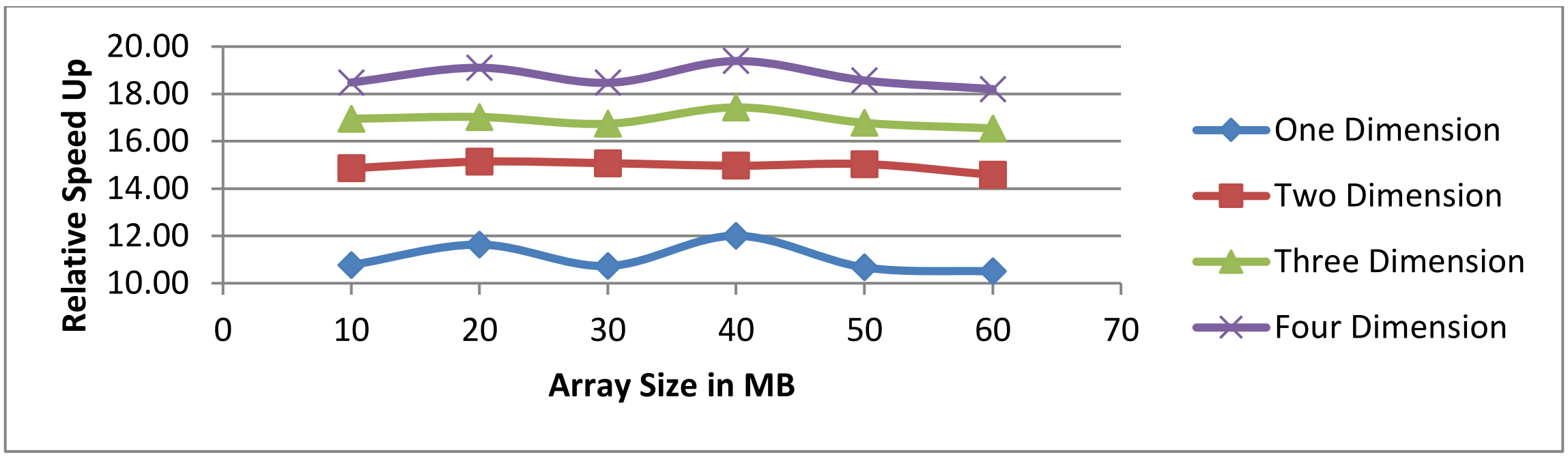

*Figure 6.9, Relative Speedup when G=P/2 using Sorted distribution for different OHHC dimensions.*

Amount of improvement when using Reversed Sorted distribution for the parallel version over the sequential version when G=P/2 and number of dimensions is 1, 2, 3, and 4 is shown in Figure 6.10. Amount of speedup is increasing when the number of processors increase, and almost steady for all array sizes. The relative speed up reached about 18% when using 4-d OHHC.

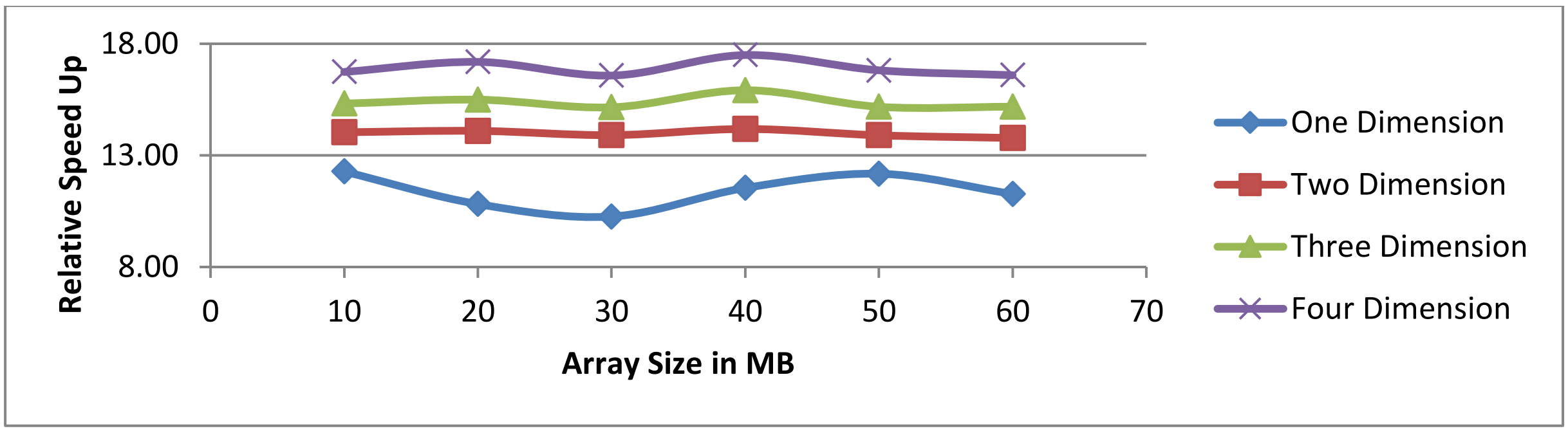

*Figure 6.10, Relative Speedup when G=P/2 using Reversed Sorted distribution for different OHHC dimensions.*

Amount of improvement when using Local distribution for the parallel version over the sequential version when G=P/2 and number of dimensions is 1, 2, 3, and 4 is shown in figure 6.11. Amount of speedup is increasing when the number of processors increase, and almost steady for all array sizes. The maximum speed up was at 4-d OHHC with only 9%, again this is due to the data distribution type.

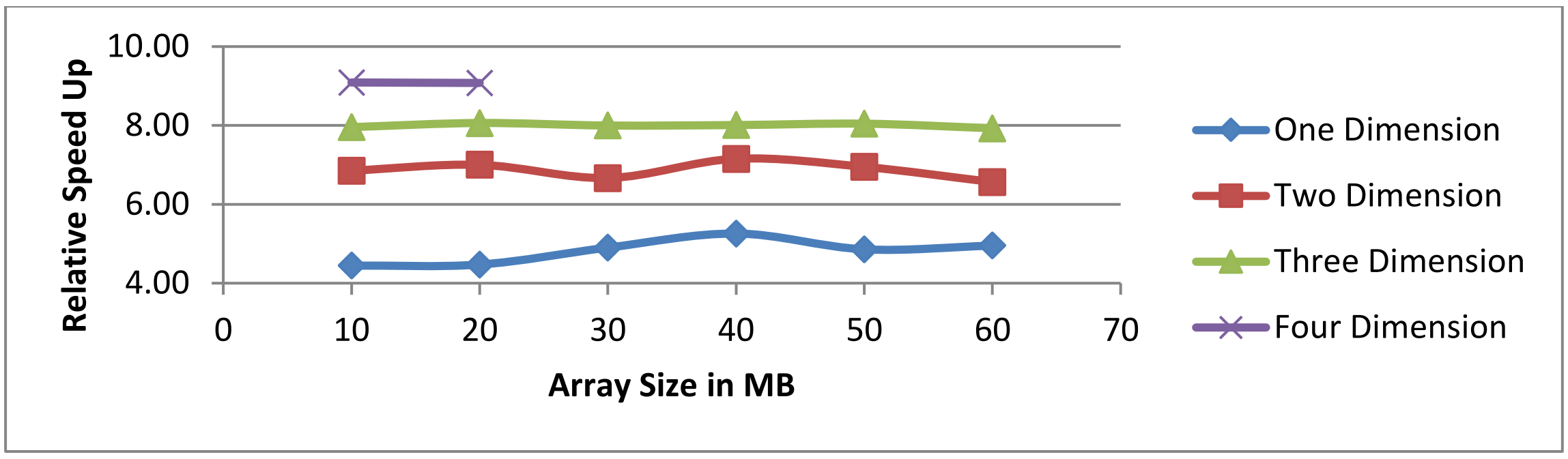

*Figure 6.11, Relative Speedup when G=P/2 using Local distribution for different OHHC dimensions.*

## 6.3. Efficiency

The proposed parallel quicksort algorithm efficiency results are presented in this section, efficiency is measured by the equation $E = T_S / P*T_P$, and represent the utilization of the engaged number of processors (threads) in the speedup as a ratio. The findings for the efficiency show that increasing the number of processors; i.e. dimensions of the OHHC, will decrease the efficiency ratio. The highest efficiency percentage where for one dimensional OHHC for all distribution types and array sizes, while the efficiency approaches zero for three and four dimensional OHHC. Moreover, we noticed that the array size has slight effect over value of the efficiency ratio. Furthermore, the highest value of efficiency where achieved when using the Sorted and Reversed Sorted distribution types. Another finding that the efficiency where almost has the same ratio when using full or half-full number of groups in the OHHC, with a slight improvement for the full OHHC over the half-full OHHC.

We will first present the efficiency results for the case where G = P, then the results for the case when G = P/2.

### 6.3.1 Efficiency results when *G = P OHHC*

Efficiency ratio were obtained for different distribution types and array sizes when number of OHHC groups is equal to number of processors in each group; i.e. number of processors in the $d_h$ HHC. The Figures of this subsection are representing the efficiency ratio for array sizes from 10MB up to 60MB of data on the x-axis and the efficiency ratio on the y-axis for different dimensions namely from 1 to 4 dimensions.

When using Random distribution, efficiency is decreasing when the number of processors increase and almost steady for all array sizes, this is expected since number of processors is increasing in polynomial ratio when increasing the OHHC dimension. However, no more than 20% efficiency was reached when using single dimension OHHC. Figure 6.12 shows efficiency ratio amount for the parallel version over the sequential version when G=P and number of dimensions is 1, 2, 3, and 4 .

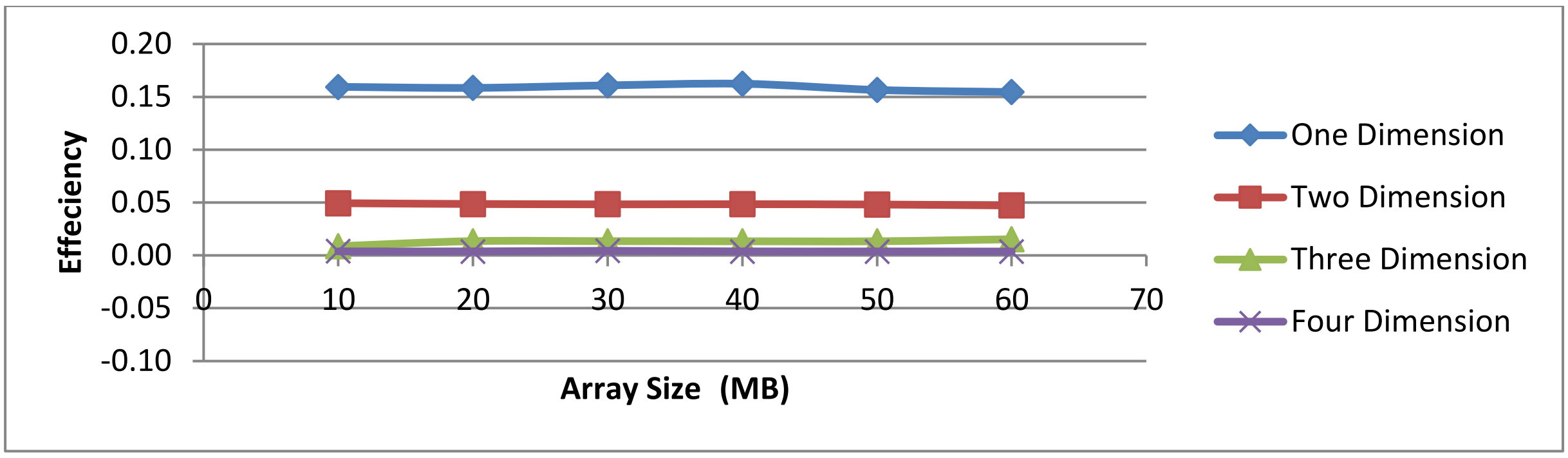

*Figure 6.12, Efficiency ratio when G=P using Random distribution for different OHHC dimensions.*

In the case of Sorted distribution, efficiency is decreasing when the number of processors increase and almost steady for all array sizes. Maximum efficiency was reached is 40% when using 1-d OHHC. Figure 6.13 shows efficiency ratio amount for the parallel version over the sequential version when G=P and number of dimensions is 1, 2, 3, and 4.

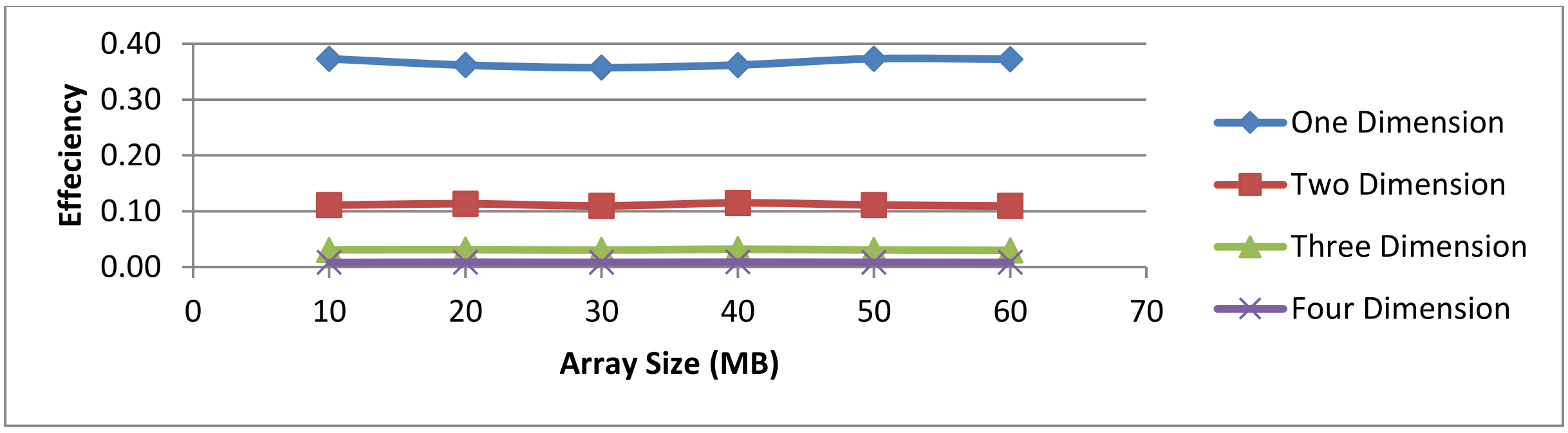

*Figure 6.13, Efficiency ratio when G=P using Sorted distribution for different OHHC dimensions.*

When using Reversed Sorted distribution, the same observation is seen as in Sorted distribution type; efficiency is decreasing when the number of processors increase and almost steady for all array sizes. The maximum reached was about 40% efficiency. Figure 6.14 shows efficiency ratio amount for the parallel version over the sequential version when G=P and number of dimensions is 1, 2, 3, and 4 .

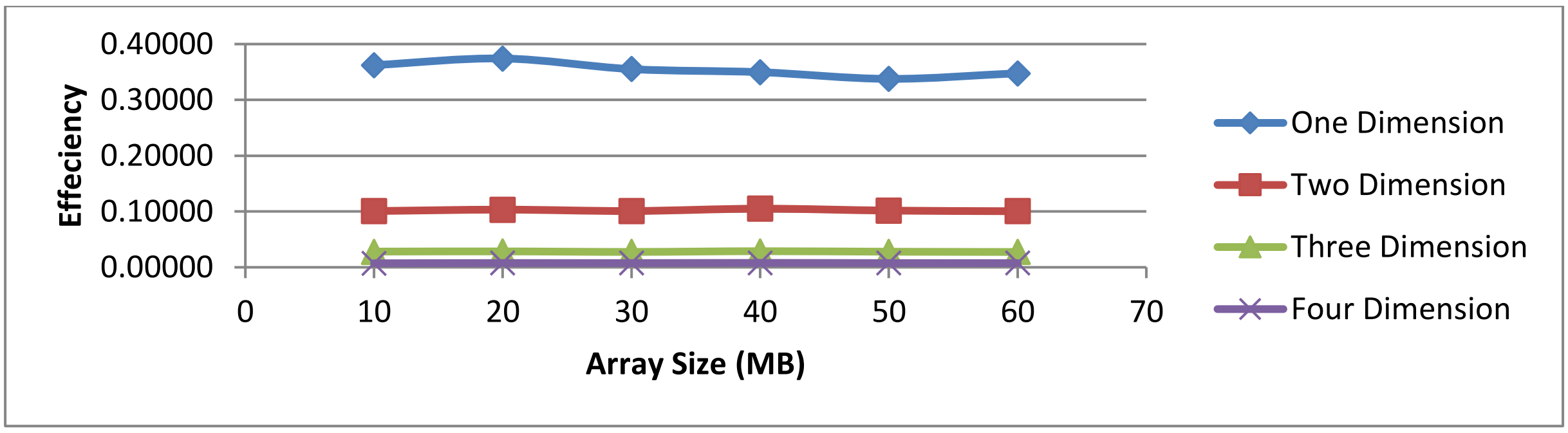

*Figure 6.14, Efficiency ratio when G=P using Reversed Sorted distribution for different OHHC dimensions.*

When using Local distribution, efficiency is decreasing when the number of processors increase and almost steady for all array sizes as the case for all distribution types but here efficiency maximum is only about 15%..

Figure 6.15 shows efficiency ratio amount for the parallel version over the sequential version when G=P and number of dimensions is 1, 2, 3, and 4 .

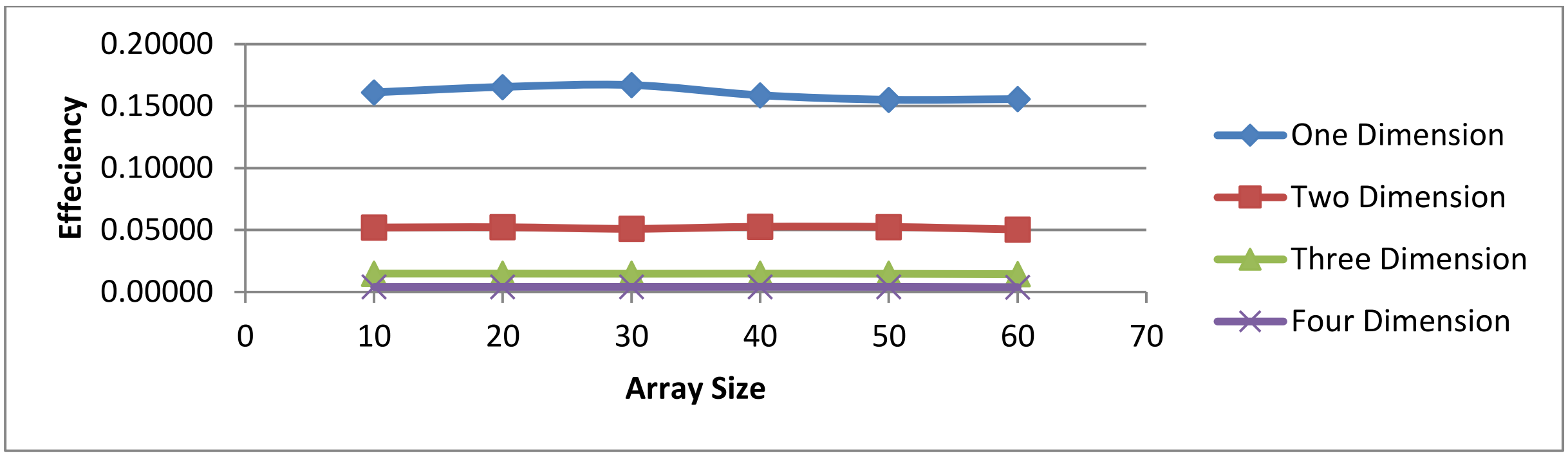

*Figure 6.15, Efficiency ratio when G=P using Local distribution for different OHHC dimensions.*

### 6.3.2 Efficiency results when G = P/2 OHHC

Efficiency ratio were obtained for different distribution types and array sizes when number of OHHC groups is equal to half the number of processors in each group; i.e. half of number of processors in the dh HHC. The Figures of this subsection are representing the efficiency ratio for array sizes from 10MB up to 60MB of data on the x-axis and the efficiency ratio on the y-axis for different dimensions namely from 1 to 4 dimensions.

When using Random distribution, efficiency is decreasing when the number of processors increase and almost steady for all array sizes, with the maximum of 15% at 1-d OHHC. Figure 6.16 shows efficiency ratio amount for the parallel version over the sequential version when G=P/2 and number of dimensions is 1, 2, 3, and 4 .

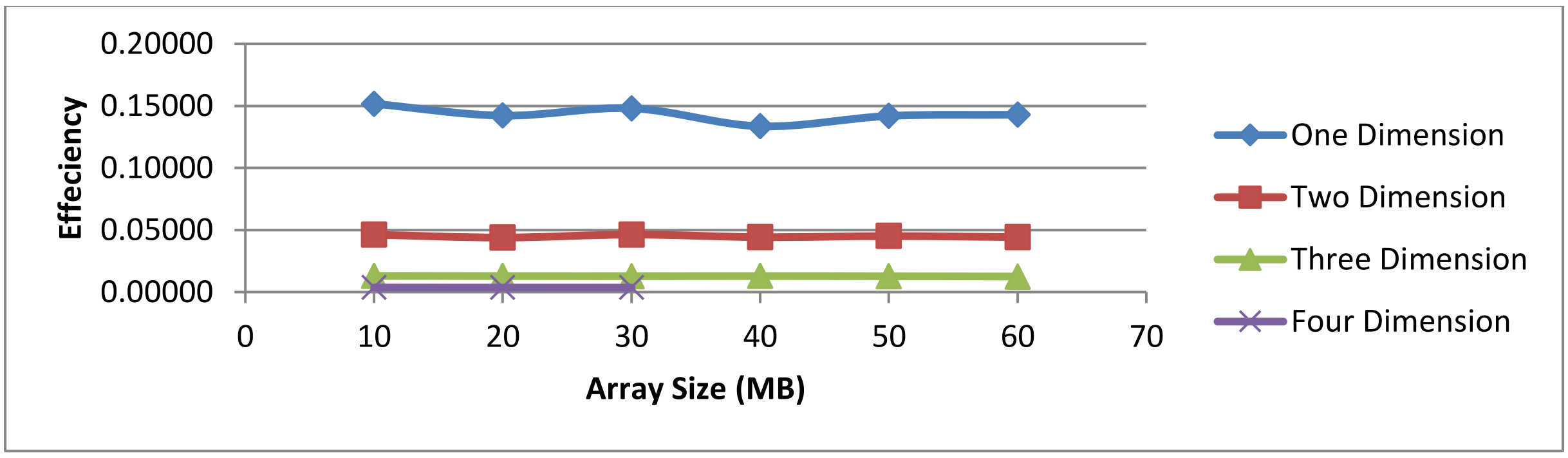

*Figure 6.16, Efficiency ratio when G=P/2 using Random distribution for different OHHC dimensions*

When using Sorted distribution, efficiency is decreasing when the number of processors increase and almost steady for all array sizes, the efficiency reached above 30% when OHHC dimension is 1. Figure 6.17 shows efficiency ratio amount for the parallel version over the sequential version when G=P/2 and number of dimensions is 1, 2, 3, and 4.

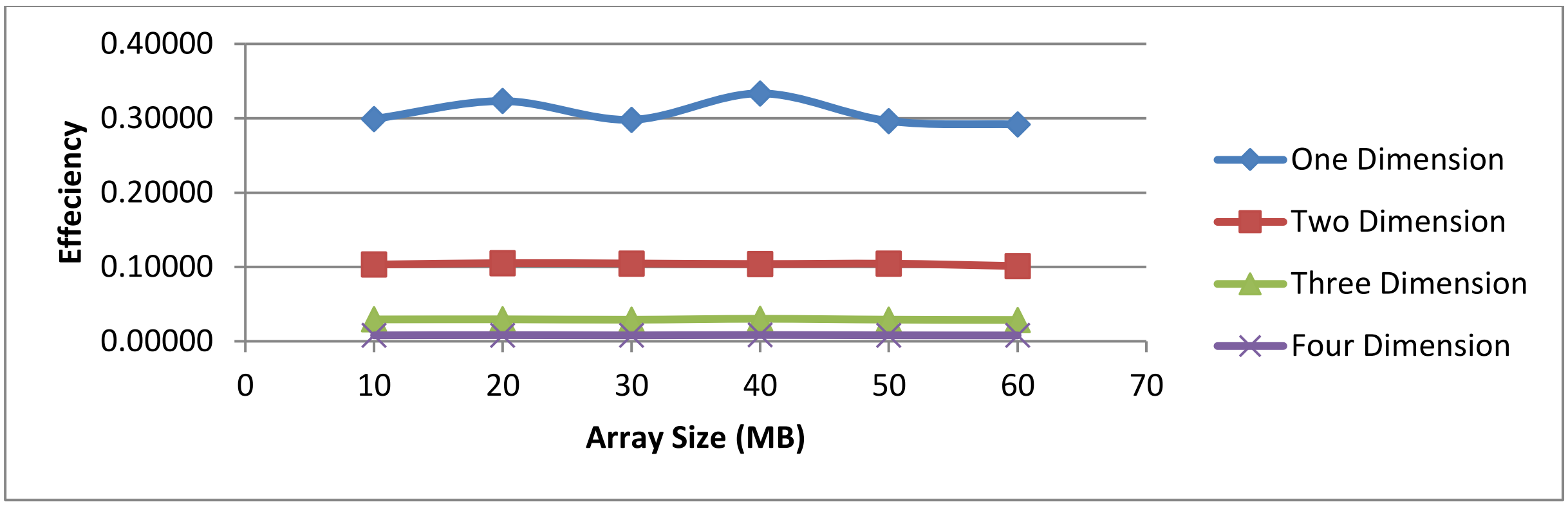

*Figure 6.17, Efficiency ratio when G=P/2 using Sorted distribution for different OHHC dimensions*

When using Reversed Sorted distribution, the same as in all distribution types; efficiency is decreasing when the number of processors increase and almost steady for all array sizes. The efficiency reached above 30% as in the case of Sorted distribution. Figure 6.18 shows efficiency ratio amount for the parallel version over the sequential version when G=P/2 and number of dimensions is 1, 2, 3, and 4 .

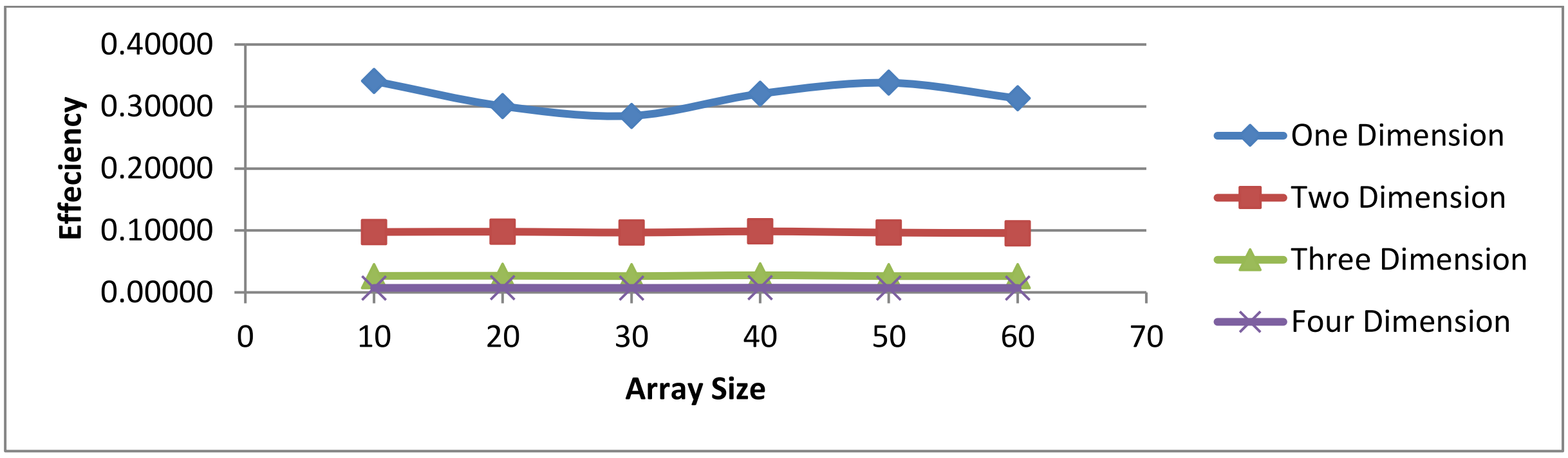

*Figure 6.18, Efficiency ratio when G=P/2 using Reversed Sorted distribution for different OHHC dimensions*

When using Local distribution, efficiency decreases when the number of processors increases and remains almost steady for all array sizes. The same observations are found in the Random distribution; efficiency remained below 15% even for single dimensional OHHC. Figure 6.19 demonstrates efficiency ratio for the parallel version over the sequential version when G=P/2 and number of dimensions is 1, 2, 3, and 4 .

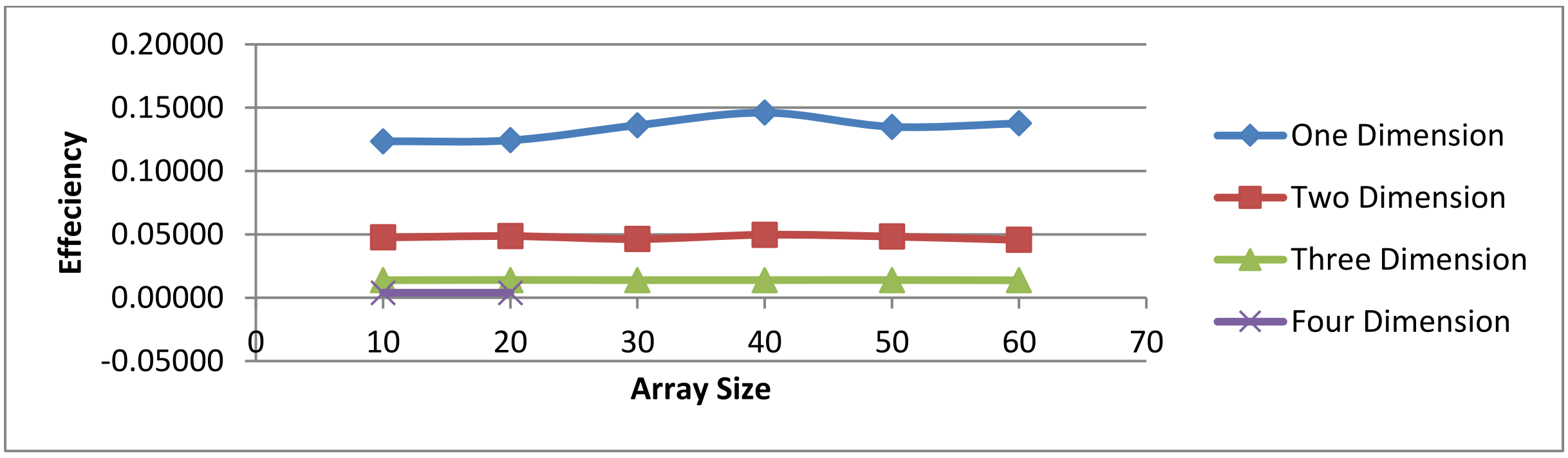

*Figure 6.19, Efficiency ratio when G=P/2 using Local distribution for different OHHC dimensions*

## Number of key comparisons

The number of key comparison was divided into three different metrics; number of recursion calls, number of iteration, and number of swaps. The results for the Random distribution type when 30MB array was sorted using OHHC dimensions from 1 to 4 is listed in figure 6.20. The figure below demonstrates that the number of recursions is steady for all dimensions and almost the swaps with a little decrease, but the number of iterations decreased significantly. These results are be reasonable since the size of the sub array were decreased but not its count.

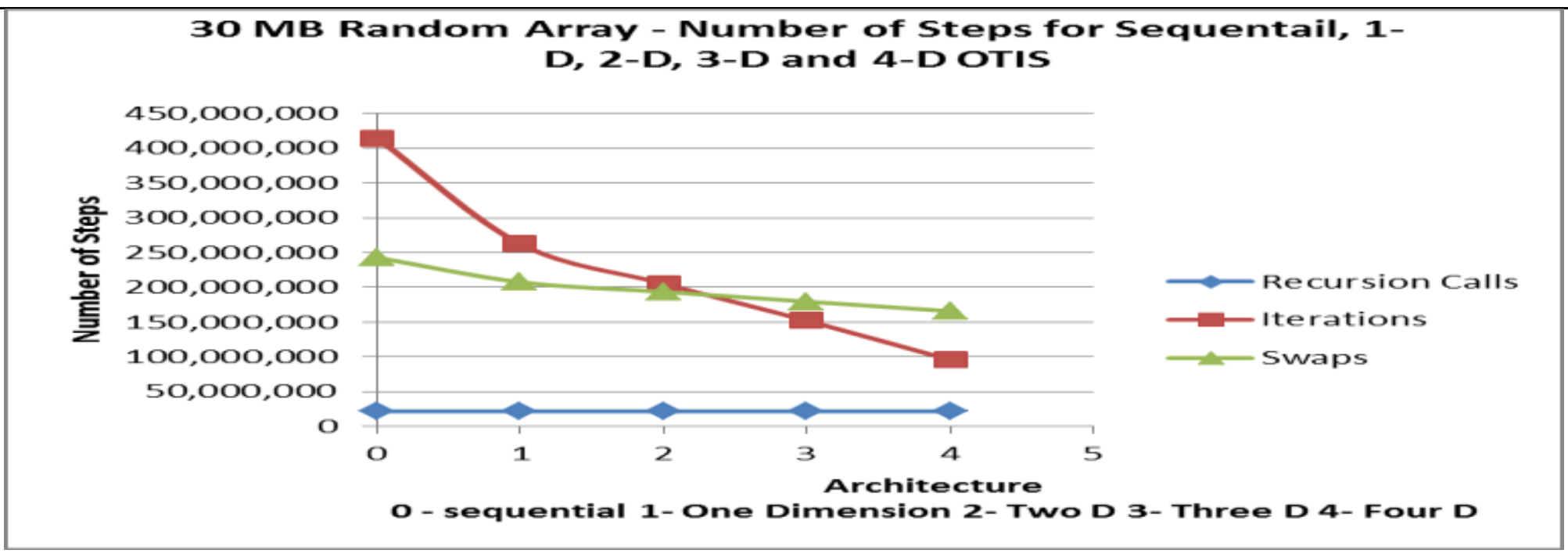

Figure 6.20, Number of recursion call, iterations, and swaps for 30MB array with random distribution using different dimension (1 to 4).

The same output from Sorted distribution type when 30MB array is presented in figure 6.21. The array was sorted using OHHC dimensions from 1 to 4. The figure demonstrates that the number of recursions is steady for all dimensions as well as the number of swaps (the values for both thus not clearly shown in the figure), but number of iterations decreased significantly. These result is reasonable since the size of the sub array were decreased but not its count.

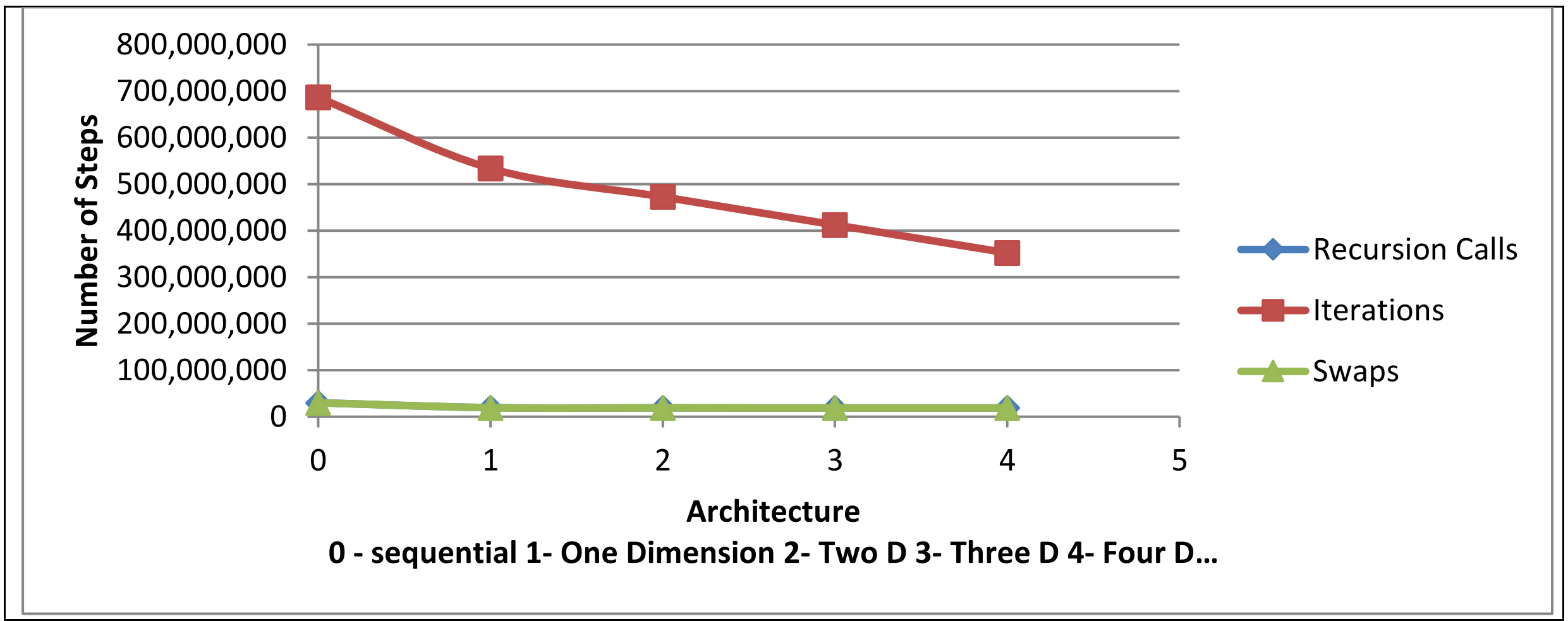

Figure 6.21, Number of recursion call, iterations, and swaps for 30MB array with Sorted distribution using different dimension (1 to 4).

A comparison between the Random and Sorted distribution types in the number of swaps is presented in figure 6.22. The big difference is due to the nature of each distribution, were in sorted type the sub arrays at the processors do not need much swapping.

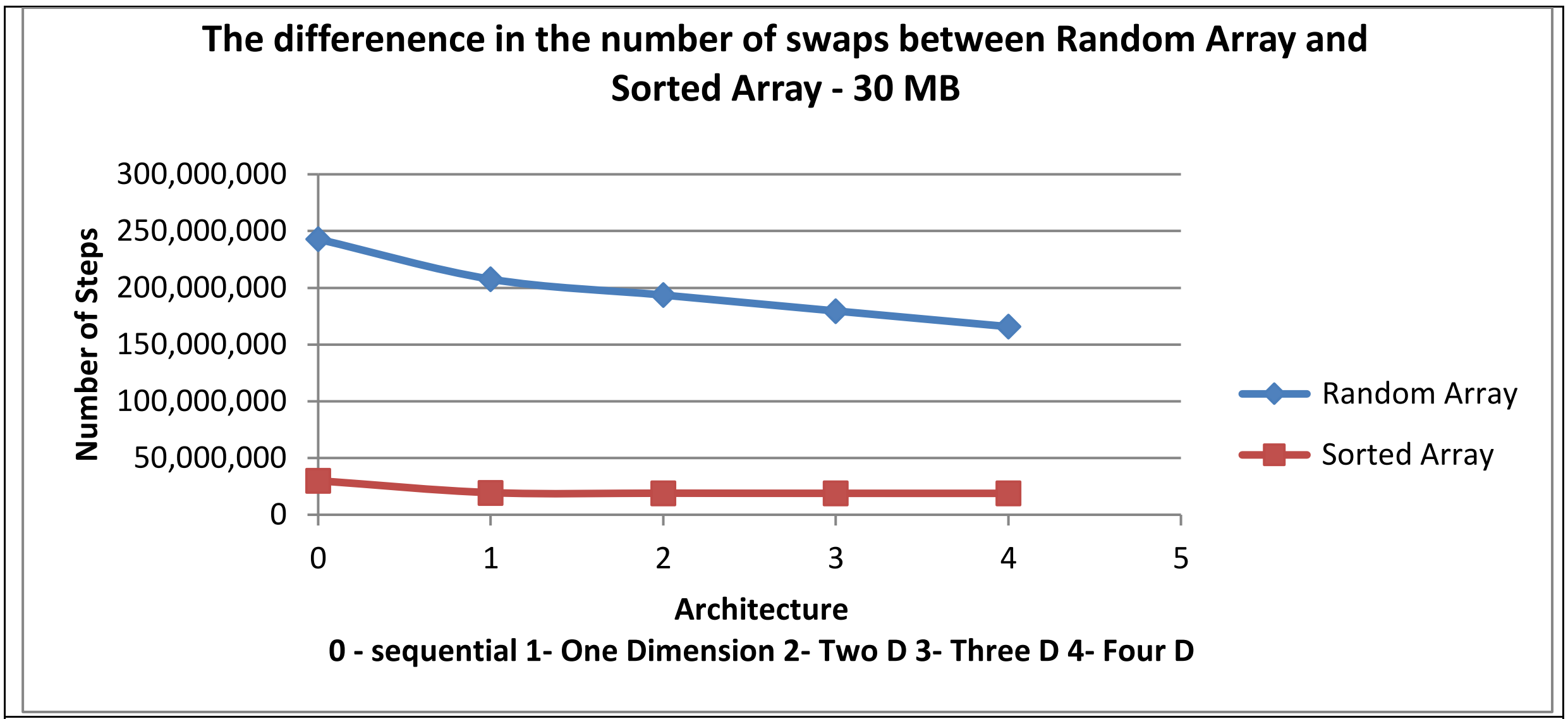

Figure 6.22, Number of swaps made by Sorted and Random distribution arrays with 30MB size using different dimension (1 to 4).

One last noticeable observation that needs to be mentioned is presented in (Figure 2.23) and Figure (2.24) below. These figures demonstrate that the pivoting process not only divided the problem, but also, this simple (O (n)) one iteration process transformed the shape of the initial problem. The applied division assisted in changing the shape of the problem by creating sub-arrays that are smaller and that has values which are more related to each other. In the initial main and large array, the member's values variance is high and ranges from small values to large values, while in the sub-arrays, the values variance is automatically minimized and the values range is also limited into a factor of thousands rather than millions.

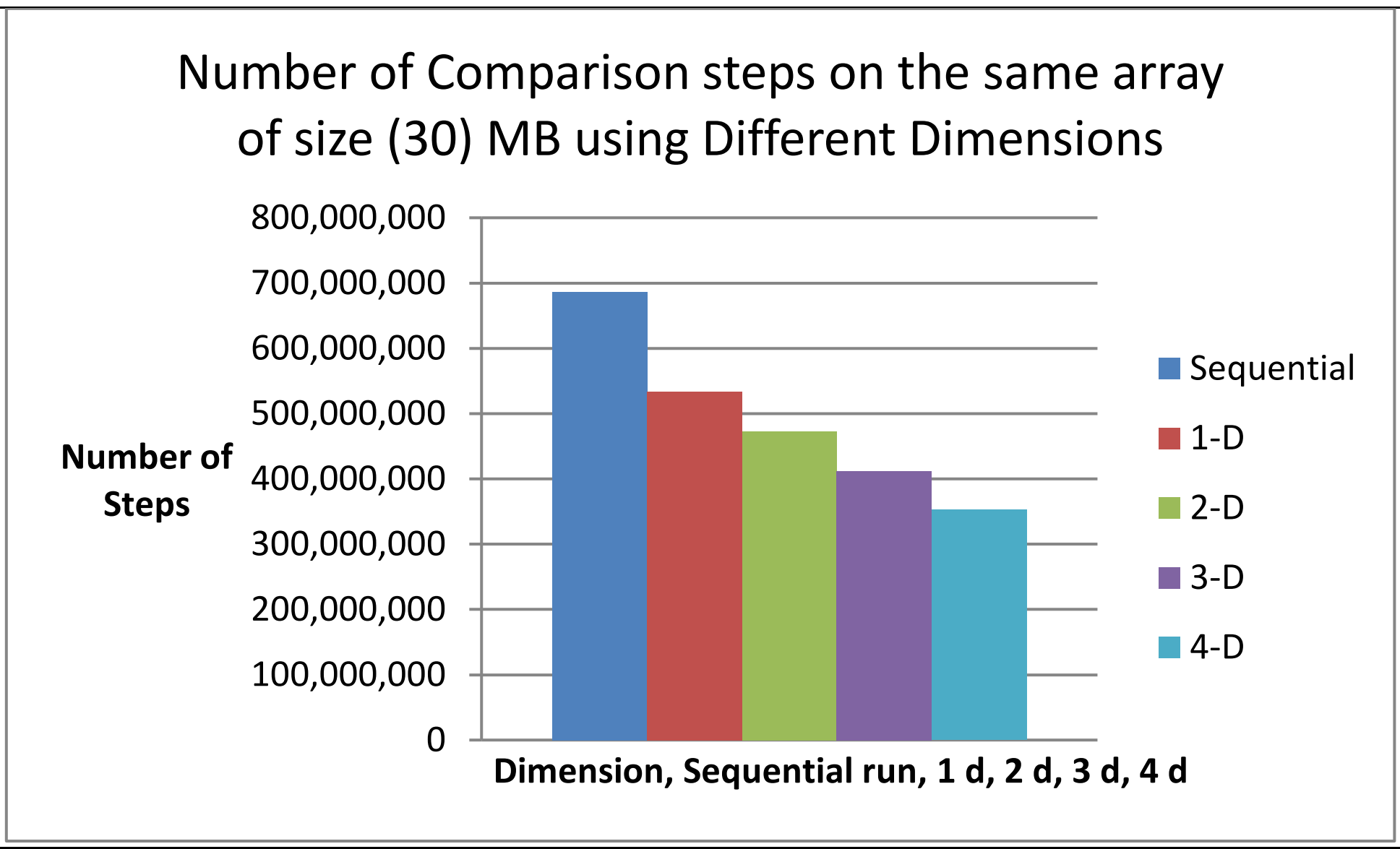

Figure 6.23, The relation between the dimension size and the number of comparison steps in the sorted array scenario, the higher the dimension, the lower the number of comparison, thus, the less computation that is required

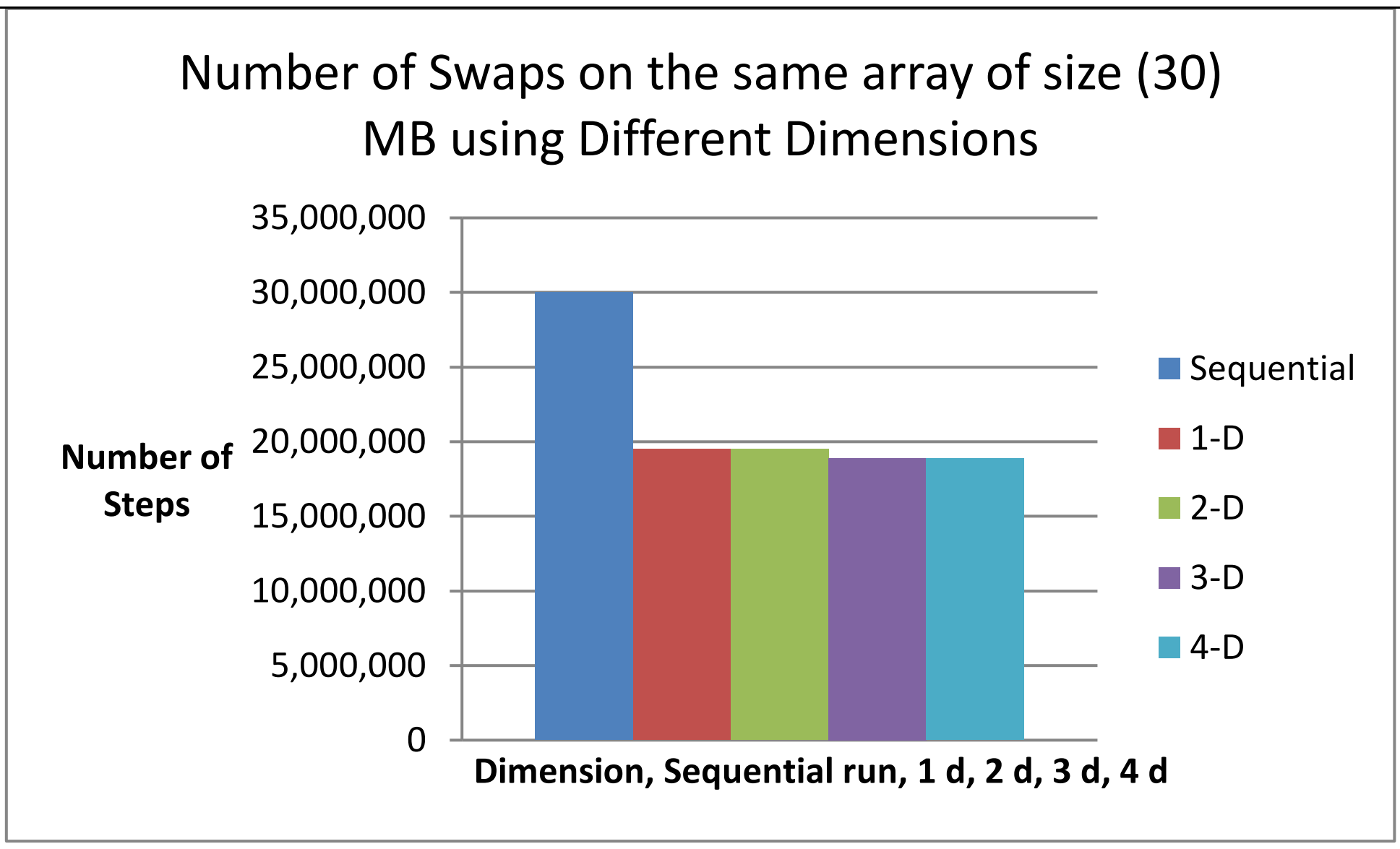

Figure 6.24, the relation between the dimension size and the number of swaps, the number of swaps is almost consistent for the sorted array scenario.

This automatic problem transformation assisted in minimizing the number of steps that were required to solve the problem, and as expected the higher the dimension is the lower the total number of computation steps that are needed. The higher dimension means larger number of pivots and more consistent smaller sub-array thus less computation, and the opposite goes true for the lower dimensions.

The bottom line is, the dividing of the problem not only makes it smaller to be solved by more resources, but also it can make it simpler and these additional resources can perform less computation.

# Conclusion

In this work we presented a parallel Quick Sort algorithm implementation that simulate the OHHC interconnection network topology by the means of multi-threading. We explored the OHHC characteristics and evaluated the algorithm analytically and by simulation. The evaluation and simulation demonstrated were analyzed, the results showed good performance of the algorithm in speedup and efficiency metrics when using both OHHC types; the full group and the half group. However, it is not realistic to decide the true performance of the OHHC interconnection network by simulation on a single CPU through multi-threading. Furthermore, the difference in the speed of the electrical and optical connections used by the OHHC was not easy to be simulated by the multi-threading and thus was not taken into consideration.

Despite of what stated above, the algorithm showed promising and encouraging results; the improvement in relative speedup reached up to 20% for both OHHC types. Efficiency improvement reached up to 40% for the OHHC full group and 30% for the OHHC half group.